\documentclass[preprint,authoryear,11pt]{elsarticle}

\usepackage{epsfig}
\usepackage{rotating}
\usepackage{lineno}
\usepackage{amssymb}
\usepackage{natbib}
\journal{Icarus}

\newcommand \dg {\mbox{$^\circ$}}
\newcommand \arcsec {\mbox{$^{\prime\prime}$}}
\begin{document}

\begin{frontmatter}



\title{Pluto's atmosphere observations with ALMA: spatially-resolved maps of CO and HCN emission and first detection of HNC}


\author{E. Lellouch$^1$,  B. Butler$^2$, R. Moreno$^1$, M. Gurwell$^3$,  P. Lavvas$^4$, T. Bertrand$^1$, T. Fouchet$^1$, D.F. Strobel$^5$,
A. Moullet$^6$
}
\address{$^1$ LESIA, Observatoire de Paris, PSL Research University, CNRS, Sorbonne Universit\'es, Université de Paris, 5 place Jules Janssen, 92195 Meudon, France; emmanuel.lellouch@obspm.fr}
\address{$^2$ National Radio Astronomy Observatory, Socorro, NM 87801, USA}
\address{$^3$ Center for Astrophysics | Harvard \& Smithsonian, Cambridge, MA 02138, USA}
\address{$^4$ GSMA, Universit\'e Reims Champagne-Ardenne, 51687 Reims Cedex 2, France}
\address{$^5$ Department of Earth and Planetary Sciences and Physics and Astronomy, Johns Hopkins University, Baltimore, MD 21218, USA}
\address{$^6$ NASA Ames Research Center Moffett Field, CA 94035, USA}


\newpage
\begin{abstract}
Pluto exhibits a tenuous, predominantly N$_2$-CH$_4$ atmosphere, with Titan-like chemistry. Previous observations with ALMA have permitted the detection of CO and HCN at 345.796 and 354.505 GHz in this atmosphere, yielding vertically resolved chemical and thermal information. We report on new observations of Pluto's atmosphere with ALMA, performed in April and July/August 2017 with two main goals: (i) obtaining spatially-resolved measurements ($\sim$0.06'' on the $\sim$0.15'' disk subtended by Pluto and its atmosphere) of CO(3-2) and HCN(4-3) (ii) targetting new chemical compounds, primarily hydrogen isocyanide (HNC) . These observations are modelled with radiative transfer codes coupled with inversion methods. The CO line shows an absorption core at beam positions within Pluto's disk, a direct signature of Pluto's cold mesosphere. Analysis of the CO line map provides tentative evidence for a non-uniform temperature field in the lower atmosphere (near 30 km), with summer pole latitudes being 7$\pm$3.5 K warmer than low latitudes. This unexpected result may point to shorter radiative timescales in the atmosphere than previously thought. The HCN emission is considerably more extended than CO, peaking at radial distances beyond Pluto limb, and providing a new method to determine Pluto's HCN vertical profile in 2017. The mean (column-averaged) location of HCN is at 690$\pm$75 km altitude, with an upper atmosphere ($>$800 km) mixing ratio of $\sim$1.8$\times$10$^{-4}$. Little or no HCN ($<$5$\times$10$^{-9}$ at 65 km) is present in the lower atmosphere, implying undersaturation of HCN there. The HCN emission appears enhanced above the low-latitude limb, but interpretation, in terms of an enhanced HCN abundance or a warmer upper atmosphere there, is uncertain. The first detection of HNC is reported, with a (7.0$\pm$2.1)$\times$10$^{12}$
cm$^{-2}$ column density, referred to Pluto surface, and a HNC / HCN ratio of 0.095$\pm$0.026, very similar to their values in Titan's atmosphere. We also obtain upper limits on CH$_3$CN ($<$2.6$\times$10$^{13}$ cm$^{-2}$) and CH$_3$CCH ($<$8.5$\times$10$^{14}$ cm$^{-2}$); the latter value is inconsistent with the reported detection of CH$_3$CCH from New Horizons. These upper limits
also point to incomplete resublimation of ice-coated aerosols in the lower atmosphere.
\end{abstract}


\end{frontmatter}


\newpage

\section{Introduction}
The exploration of the Pluto system by New Horizons has provided a once-in-a-lifetime view of Pluto's atmosphere, revealing its thermal structure, gas composition,
haze, dynamics, interactions with the surface, and escape characteristics \citep[see reviews in][and references therein]{stern18,gladstone19,young21,forget21}.
Pluto's atmosphere shares fundamental characteristics, both with Mars' (e.g. regarding the role of volatile cycles), and with Titan's (e.g. a rich nitrogen-methane
photochemistry with an extended haze). Although the New Horizons flyby represents a single snapshot of a time-variable object, the overwhelming amount of relevant atmospheric, surface composition and geomorphology data it yielded has led to a basic understanding of Pluto's atmosphere and climate. This can now be used to predict the 
seasonal evolution of the body and which represents a benchmark for the study of similar objects, notably the surface-atmosphere system of Triton and possibly
of other volatile-rich KBOs \citep{young20}. Ground-based observations, while in most cases not rivaling with spaceborne exploration for such a small and distant world, have provided additional and often unique contributions to the characterization 
of Pluto's atmosphere, enabling in particular the monitoring of the surface pressure with time from stellar occultations \citep[][and references therein]{meza19,arimatsu20}. Ground-based atmospheric spectroscopy prior to New Horizons has also secured the first measurements of CH$_4$ \citep{young97,lellouch09} and the detection of two other gases
(CO and HCN) not observable by New Horizons \citep{lellouch11,lellouch17}. Disk-averaged ALMA observations \citep[][hereafter Paper I]{lellouch17} of CO(3-2) and HCN(4-3) determined the mean abundance of these gases, and established the existence of an unexpectedly cold ($\sim$70 K) upper atmosphere, in agreement with New Horizons findings \citep{gladstone16,young18}. This low temperature throttles atmospheric escape to inconsequential values, but a conclusive explanation for the cold temperatures remains to be found  \citep{strobel17,zhang17}.

Except for the exquisitely detailed imaging of the haze, spatially-resolved sounding on Pluto's atmosphere remains poor. The two most diagnostic
observations from New Horizons were the REX radio-occultations, providing pressure-temperature-altitude profiles in the lowest $\sim$120 km \citep{hinson17}, and the UV solar occultations (Alice) returning N$_2$, CH$_4$ and other hydrocarbons, and haze profiles in various altitude ranges altogether covering the 0-1200 km range \citep{young18}.
These datasets were obtained only at two low-latitude locations (ingress: 15.5\dg~S (Alice) -- 17\dg~S (REX); egress: 16.5\dg~N (Alice) -- 15.1\dg~N (REX)). Except for differences in the temperature structure of the lower atmosphere ($<$ 30 km) associated with surface boundary layer and topographic effects \citep{hinson17}, these measurements indicated very similar temperature, gas and haze profiles at ingress and egress. Additional (but lower quality, compared to solar occultation) UV data were obtained from a stellar occultation and a stellar appulse, extending the sounded latitudes to 40\dg~S \citep{kammer20}, but polar latitudes were not sampled. 

The ALMA array yields sufficient spatial resolution -- up to 0.012'' at 345 GHz for the most extended configurations -- to comfortably resolve Pluto's solid body (0.10'' as seen from the Earth) and its atmosphere. In practice, such ultra-high resolutions are not achievable due to the weaknesses of the Pluto atmospheric signals (60-100 mJy, integrated) and the concomittant need for high spectral resolution. However, intermediate array configurations of several 10's of milli-arcseconds (mas) do allow resolution of Pluto's atmosphere, while providing sufficient sensitivity. A specific advantage is that currently Pluto's North pole is well positioned for viewing from Earth (sub-Earth latitude on Pluto $\sim$ 50-57\dg~N over 2015-2020) enabling the investigation of equatorial-to-pole variations in the temperature and composition fields. Furthermore, the ALMA sensitivity offers the opportunity to search for additional species,
potentially complementing the known molecular inventory in Pluto's atmosphere. We hereby present new ALMA observations of Pluto's atmosphere, performed in 2017, whose highlights
are (i) the first mapping of CO and HCN mm emission, and (ii) the first detection of hydrogen isocyanide (HNC). A preliminary assessment of the data and initial results have been
given in \citet{lellouch18}.

\section{Observations and data reduction}\label{sec:obs}


   We obtained a new set of observations of the Pluto-Charon system with
the 12-m array of the Large Millimeter Array (ALMA), under the ALMA
project 2016.1.00426.S. The goals of the project were (i) to acquire
spatially-resolved data of the CO and HCN line emissions detected in
Paper I (ii) to search for new additional atmospheric compounds, in
particular hydrogen isocyanide (HNC).  Observations were conducted on
April 27, July 30, August 4 and August 7, 2017, using two different
frequency set-ups.  The ``HNC setup'' (used on April 27) targeted primarily the HNC(3-2)
line at 271.981 GHz, while also covering frequency ranges around
H$^{13}$CN(3-2) (259.012 GHz), HC$^{15}$N(3-2) (258.157 GHz), CH$_3$CN J
= 13 multiplet (257.448 GHz) and CH$_3$CCH J = 14 multiplet (256.258
GHz).  The spectral resolution was 122 kHz for HNC and 244 kHz for other
lines.  The ``CO/HCN setup'' (used on the three other dates) covered the CO(3-2) and HCN(4-3) lines at
345.796 and 354.505 GHz, respectively, while also covering frequency
ranges around HC$_3$N(38-37) and (39-38) at 345.609 and 354.697 GHz.
The spectral resolution was 122 kHz for all lines.  Both setups also
included continuum spectral windows, with 1.875 GHz bandwidth and 0.976 MHz resolution.  There
were two such ranges in the CO/HCN setup, and one in the HNC setup; the
HNC continuum window also included the HC$_3$N(30-29) line at 272.885 GHz.

\begin{table}
\caption{ \label{obs-tab}
   Observing dates and geometry.
   }
\centering
\footnotesize
\begin {tabular}{ccccccc}
\hline\hline
\noalign{\vspace{3pt}}
Date & Time range (UTC) & Geocentric & Sub-Earth &
Pluto-Charon  & Set-up & Synthesized beam\\
      &                  &       Distance (AU)         & Longitude$^a$ &
separation (arcsec) &  \\
\noalign{\vspace{3pt}}
\hline
\noalign{\vspace{2pt}}
2017-Apr-27 & 07:21--08:12 & 32.994 & 58-56 & 0.77 & HNC & 0.62'' x 0.45'', PA$^b$ = 89\dg \\
2017-July-30 & 05:42--07:04 & 32.418 & 165-162 & 0.69 & CO/HCN  & 0.091'' x 0.051'', PA=-76\dg \\
2017-Aug-04 & 01:48--03:17 & 32.452 & 253-249 & 0.81 & CO/HCN & 0.074'' x 0.053'', PA=-74\dg\\
2017-Aug-07 & 01:06--02:31 & 32.477 & 85-82 & 0.83 & CO/HCN & 0.068'' x 0.053'', PA=-81\dg\\
\noalign{\vspace{2pt}}
\hline\hline
\multicolumn{7}{l}{\footnotesize $^a$ Sub-observer East longitude at mid-point. We adopt the new IAU convention for definition of the North Pole} \\
\multicolumn{7}{l}{\footnotesize \citep{buie97,zangari15} with current spring in the northern hemisphere. Zero longitude on Pluto is the }\\ 
\multicolumn{7}{l}{\footnotesize sub-Charon point and the sub-observer point longitude decreases with time.} \\
\multicolumn{7}{l}{\footnotesize $^b$ Beam position angle, measured from celestial north. For the combined CO/HCN data, the synthesized beam is}\\
\multicolumn{7}{l}{\footnotesize  0.076'' x 0.052'', PA=-76\dg}\\
\end{tabular}
\end{table}

   Because the prime goal of the HNC setup observations was detection,
we used a low spatial resolution (ALMA configuration C40.3, with maximum
baseline $\sim$460 m and resolution $\sim$0.58$\arcsec$), sufficient to
separate Pluto from Charon, while ensuring the entire flux density from
Pluto would be measured.  For the CO/HCN setup, the prime goal
was to resolve Pluto.  As the apparent size of Pluto was
$\sim$0.10$\arcsec$  ($\sim$0.15$\arcsec$ at 600 km altitude),
we observed that setup in the more extended ALMA
configuration C40.7, with maximum baselines $\sim$3700 m and resolution
$\sim$0.06$\arcsec$.  Details on the observational parameters for each date are given
in Table \ref{obs-tab}. The HNC setup observation had a duration of 50 min, while
each CO/HCN observation had a duration of 85 min.  For all observations,
J1924-2914 was used to calibrate the bandpass and flux density scale,
while J1911-2006 was used to calibrate temporal fluctuations in complex
gain (amplitude and phase, both instrumental and atmospheric).  An
additional source, J1923-2104, was included as a ``check source''.  ALMA
includes these sources to allow for an atmospheric decorrelation
correction, if needed.  However, as described below, there was enough
flux density to self-calibrate the data so this correction was not
necessary.


  Data reduction followed the initial steps outlined in Paper I.  This
included calibration and velocity corrections on the spectral
visibilities, performed with the CASA pipeline (Muders et al., 2014).
Further data processing, including self-calibration, visibility fitting,
deconvolution, and imaging were then performed in both the
AIPS\footnote{http://www.aips.nrao.edu} and
GILDAS\footnote{https://www.iram.fr/IRAMFR/GILDAS} post-processing
packages.


   For all four observations, we then imaged the continuum spectral windows (combining the two, for the CO/HCN
setup), and used that image to perform self-calibration to correct for
atmospheric phase fluctuations, also applying the determined solutions
to the spectral line spectral windows.  We then fitted the
self-calibrated visibilities to determine the continuum brightness
temperature of Pluto and Charon, and found significant fluctuation
between the four days, for both bodies.  We
believe that this is because either the flux density scale calibrator
(J1924-2914) was fluctuating, or the application of its flux density to
J1911-2006 (and thence to our target data) was incorrect.  Similar
offsets have been found when using ALMA grid calibrators to set the flux
density scale in other observations \citep{francis20}.   The 2015 observations (Paper I) did not suffer
from this issue as they used Titan as the flux density scale
calibrator, which has a very good emission model \citep{butler12}.  Because of
this, we fixed the flux density scale such that the brightness
temperature of Pluto was 33 K, as was found in our 2015 observations
(Paper I; Butler et al. 2020). This brightness temperature results in a continuum flux density level of 17.5
mJy at CO(3-2), 18.3 mJy at HCN(4-3), and 11.1 mJy at HNC(3-2).  While the continuum
thermal emission of Pluto and Charon may vary with the system's
heliocentric distance $r_h$, a simple scaling of temperatures as
$r^{-0.5}_h$ would lead to an expected temperature decrease of only
0.7\% (and the latter could be offset by the increase of the
subearth/subsolar latitude, causing the fraction of (more illuminated)
high-latitudes to increase over 2015-2017).  These steps were all performed in
AIPS.

   \begin{figure}
   \includegraphics[angle=0,width=18cm]{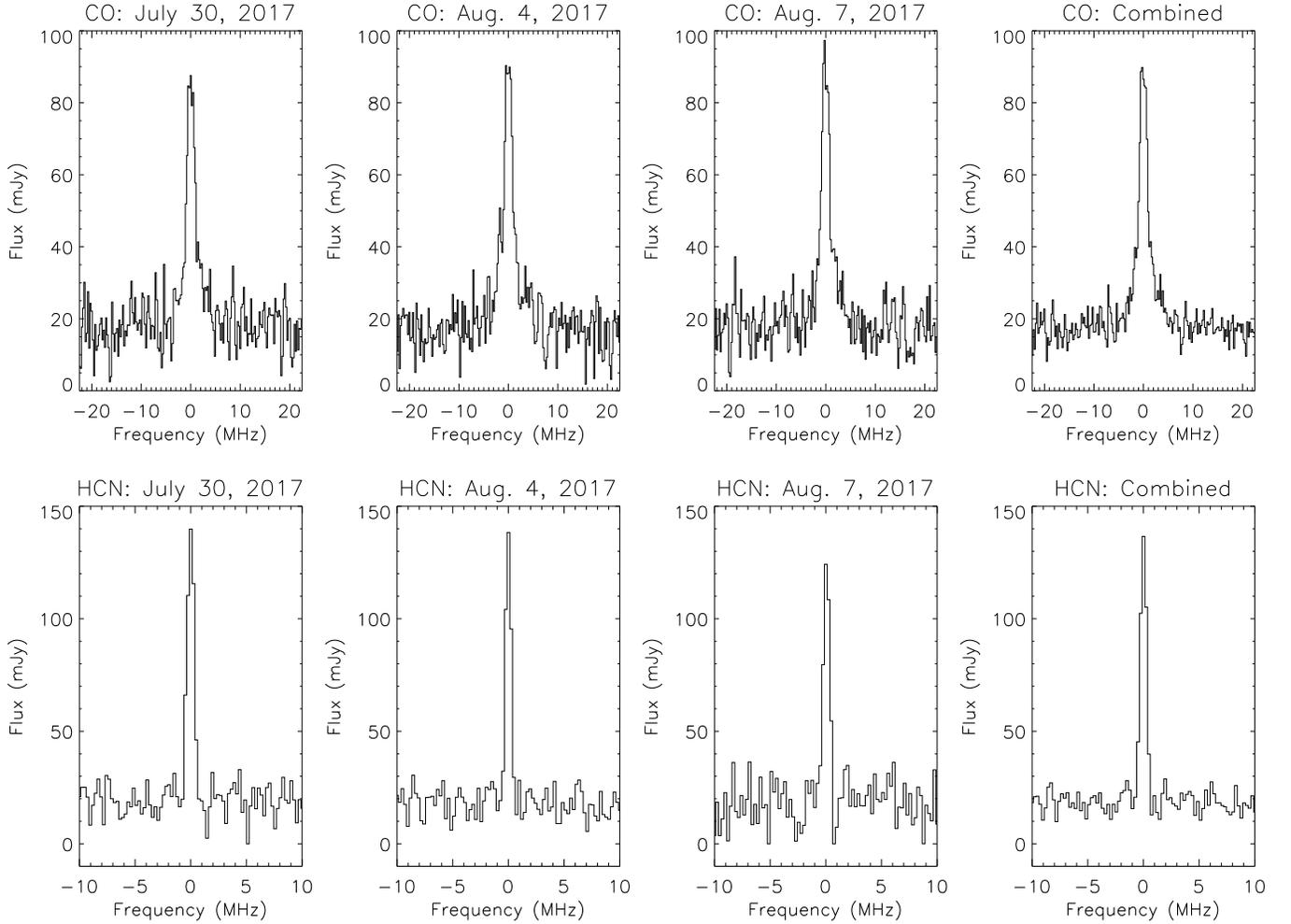}
   \caption{Overview of CO(3-2) (top) and HCN(4-3) (bottom) zero-spacing spectra on the three individual days, and combined. }%
   \label{zerospacing}
    \end{figure}

   For the CO/HCN setup observations, we next fitted the visibilities in
the spectral line windows on a per-channel basis, in order to determine
the best value of the total flux density (a.k.a. zero spacing flux) as a function of channel.  For
this fitting, we excluded the Charon contribution and fitted the visibilities over a maximum u-v range of 1000 k$\lambda$, forcing Pluto to be
at phase center. Pluto's continuum was described as emission from a 0.1009" disk with some limb darkening (0.1 exponent). 
For the lines, we prescribed channel-dependent values of the emission
size based on modeling of the image cube data (see below).  For the CO
spectral window, the specified emission size was 1.01, 1.05, and 1.40
times the diameter of Pluto, respectively, at 5 MHz, 1 MHz, and 0 MHz
from line center (those values were interpolated for intermediate
frequency offsets).  For HCN, the specified emission size was 1.00,
1.40, and 1.78 times the diameter of Pluto, respectively, at 0.6 MHz,
0.5 MHz, and 0.0 MHz from the line center.  This was done in AIPS, but
was also repeated in GILDAS, leaving the fitted diameter as a free
parameter; this yielded consistent results.  We did this fitting separately
for each date, and also with the data combined over all
the three observations.  The corresponding (total flux density) HCN and CO
spectra for the individual days, and for data combined over the three
days are shown in Fig.~\ref{zerospacing}. Within noise level, the CO and HCN lines are the same for the three individual dates, although the HCN line contrast observed
on August 7 is lower (105$\pm$9 mJy) than on July 30 and August 4 (120$\pm$7 and 121$\pm$6 mJy) at the 1-$\sigma$ level. In what follows, only the combined CO/HCN data were
imaged and analyzed. The synthesized beam for this combined data
set is 0.076'' x 0.052''.

   Finally we produced image cubes for all of the spectral
windows, for the combined CO/HCN data, and the HNC
observation.  The image cube for the CO/HCN data was made with
$0.01\arcsec$ pixels, in order to subsample the synthesized beam
sufficiently 
While our resulting resolution does not heavily resolve Pluto, the large
subearth latitude ($\beta = 53.5^\circ$) at the time of the observation
and the preferentially East-West orientation of the beam (polar angle PA
= –76$^\circ$) are favorable to separate the high and low latitudes, as
illustrated in Figure 2.  For the HNC setup, which did not spatially
resolve Pluto (resolution $\sim$0.62$\arcsec$ x $\sim$0.45$\arcsec$), the image cube was made with $0.1\arcsec$ pixels,
and the spectra  were simply extracted at image center.  The spectral channels were binned by a factor 4 to enhance S/N,
yielding a spectral resolution of 488 kHz for HNC and 976 kHz for
H$^{13}$CN, HC$^{15}$N, CH$_3$CN and CH$_3$CCH.



\section{CO and HCN: data presentation}
\label{COandHCNdata}
\begin{figure}
   \includegraphics[angle=90,width=15cm]{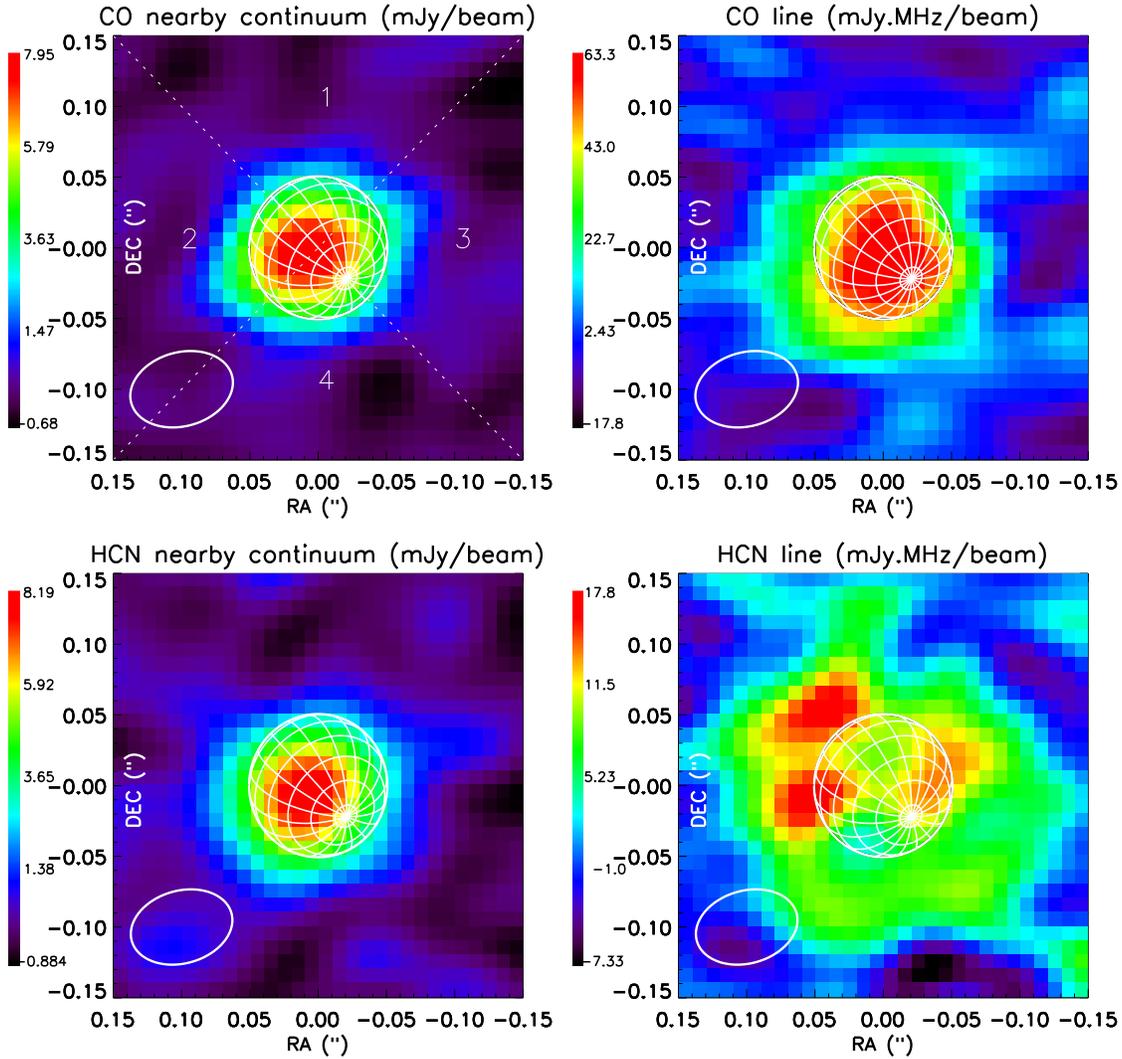}
   \caption{Images of continuum (mJy/beam) and CO and HCN line (mJy$\cdot$MHz/beam) emission from Pluto. The CO and HCN emissions
are integrated over $\pm$5 MHz and $\pm$1 MHz, respectively from line center. The synthesized beam is shown as a white oval and the Pluto globe -- with a 20\dg\ lat/lon grid, North Pole visible -- is indicated. Given Pluto's subsolar latitude of 53.5\dg N, only latitudes higher than $\sim$35 S are visible, and in practice Pluto's southern latitudes do not contribute much to the observed signal. The white dashed lines in the first panel delineate the quadrants that are being used in producing "Low/High latitudes" and "Morning/Evening" averages (see text). }%
   \label{Fig_COHCN_maps}
    \end{figure}

Figure~\ref{Fig_COHCN_maps} shows images of the continuum and CO(3-2) and HCN(4-3) line emission from Pluto. The CO and HCN emissions, expressed in mJy$\cdot$MHz/beam, are integrated 
over $\pm$5 MHz and $\pm$1 MHz, respectively, from line center, and the continuum (in mJy/beam) is taken in the immediate vicinity of the lines ($\pm$10-40 MHz from line center).
Immediately apparent is the fact that the CO emission is somewhat more extended than the continuum, while the HCN emission is considerably wider. This is qualitatively
consistent with results from Paper I which -- based on radiative-transfer analysis of spatially unresolved line profiles -- showed that the CO(3-2) line probes up to
$\sim$400 km in Pluto's atmosphere, while the HCN emission in the line core mostly originates from the 500-1000 km altitude range. Fig.~\ref{Fig_COHCN_maps} further shows
that 
the CO emission appears to be enhanced at high Northern latitudes compared to Equatorial latitudes, and the opposite is generally seen for HCN.

   \begin{figure}
   \includegraphics[angle=90,width=15cm]{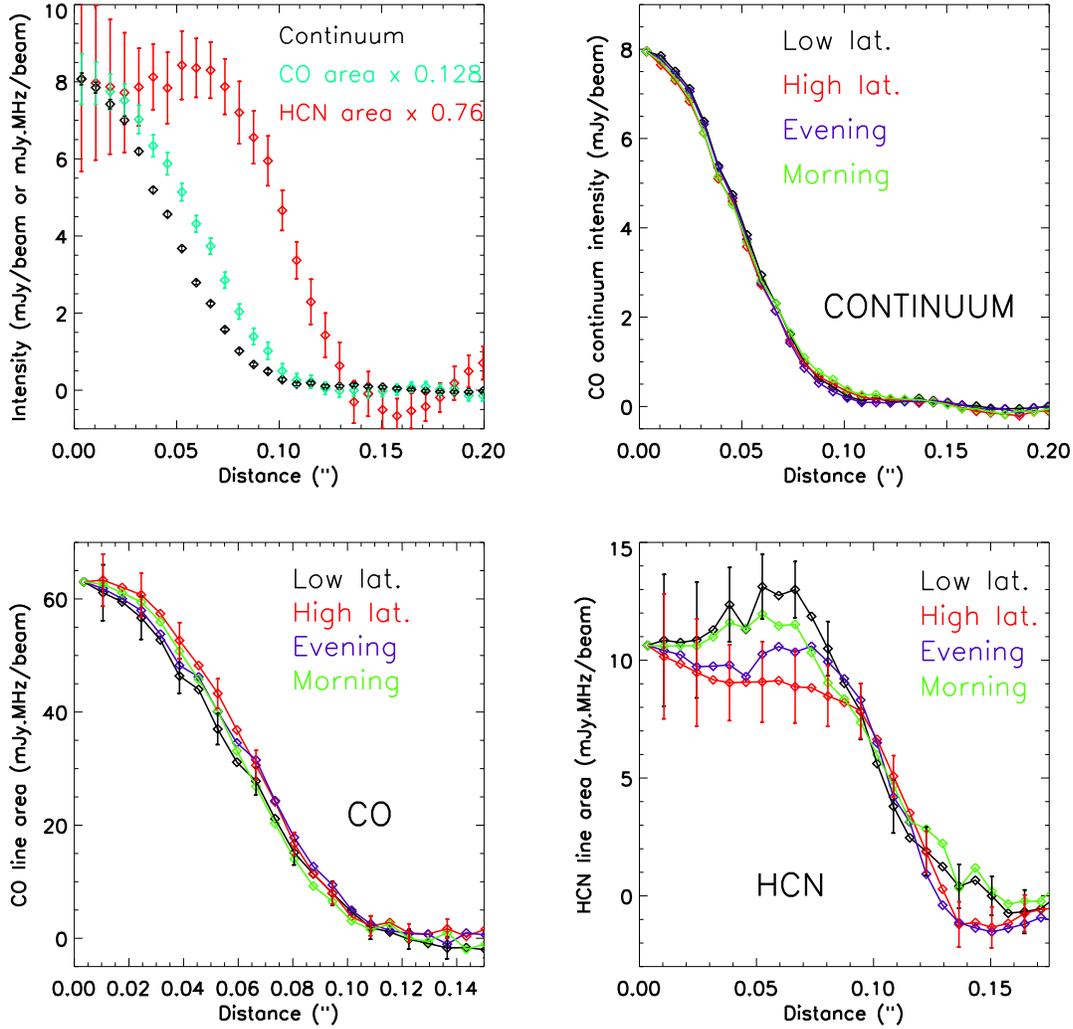}
   \caption{Radial profiles of continuum (mJy/beam) and CO and HCN line emission (mJy$\cdot$MHz/beam) from Pluto. (Top left panel): data from Fig.~\ref{Fig_COHCN_maps}
are integrated over radial bins of 0.007". To ease comparisons, the CO and HCN emissions are multiplied by constant factors (0.128 and  0.76 respectively) so that all profiles
coincide in the first bin centered at r~=~0.0035". In the other three panels, the continuum, CO and HCN emissions are plotted after being split in two pieces, 
either as "Low / High latitudes" or as "Morning / Evening" (see text for details). Pluto's mean apparent radius for the observation dates is 0.0505''.}
    \label{Fig_radialprofiles}
   \end{figure}

    \begin{figure*}
   \includegraphics[angle=90,width=8cm]{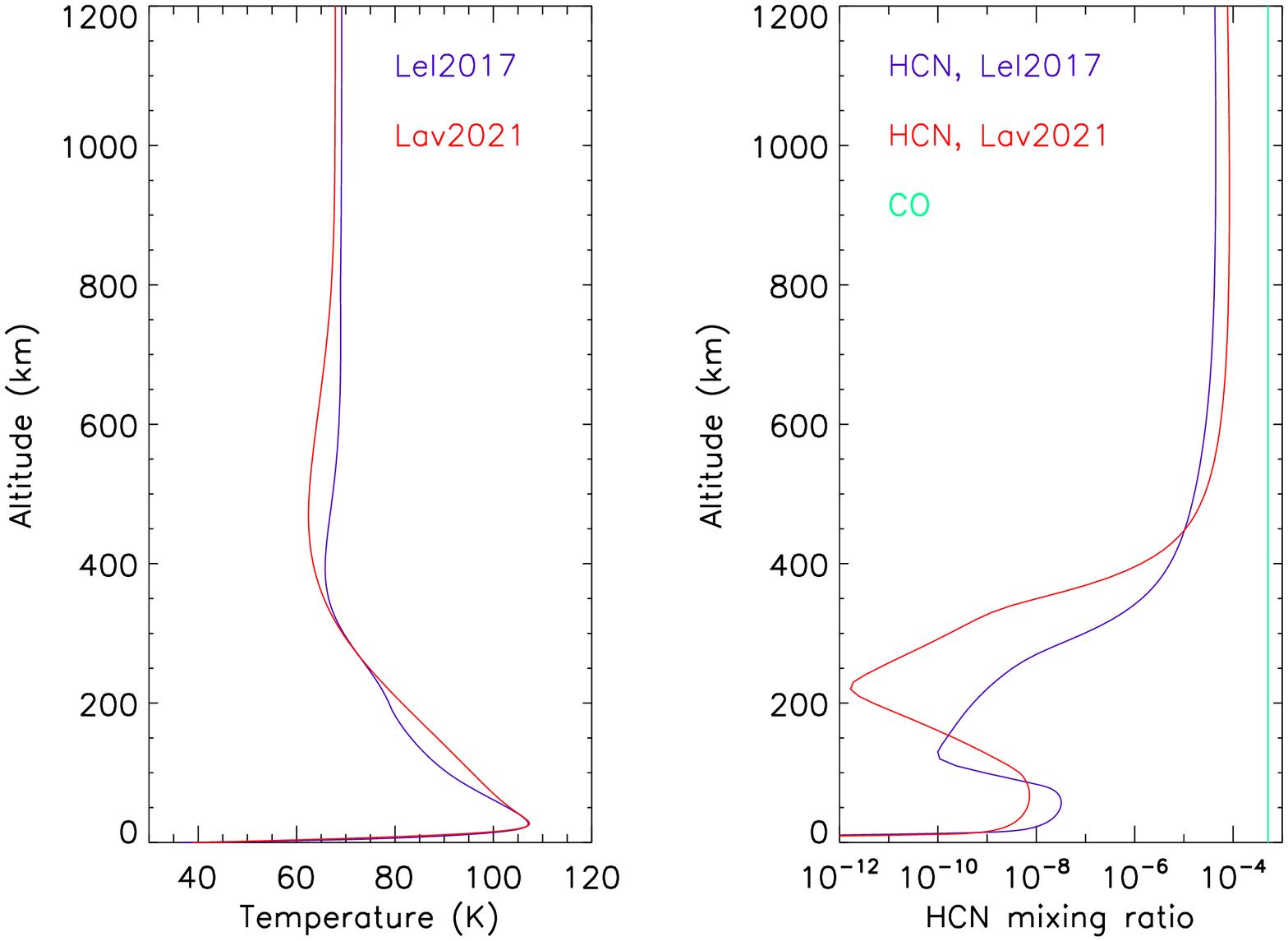}
\hspace*{1cm}
   \includegraphics[angle=90,width=8.15cm]{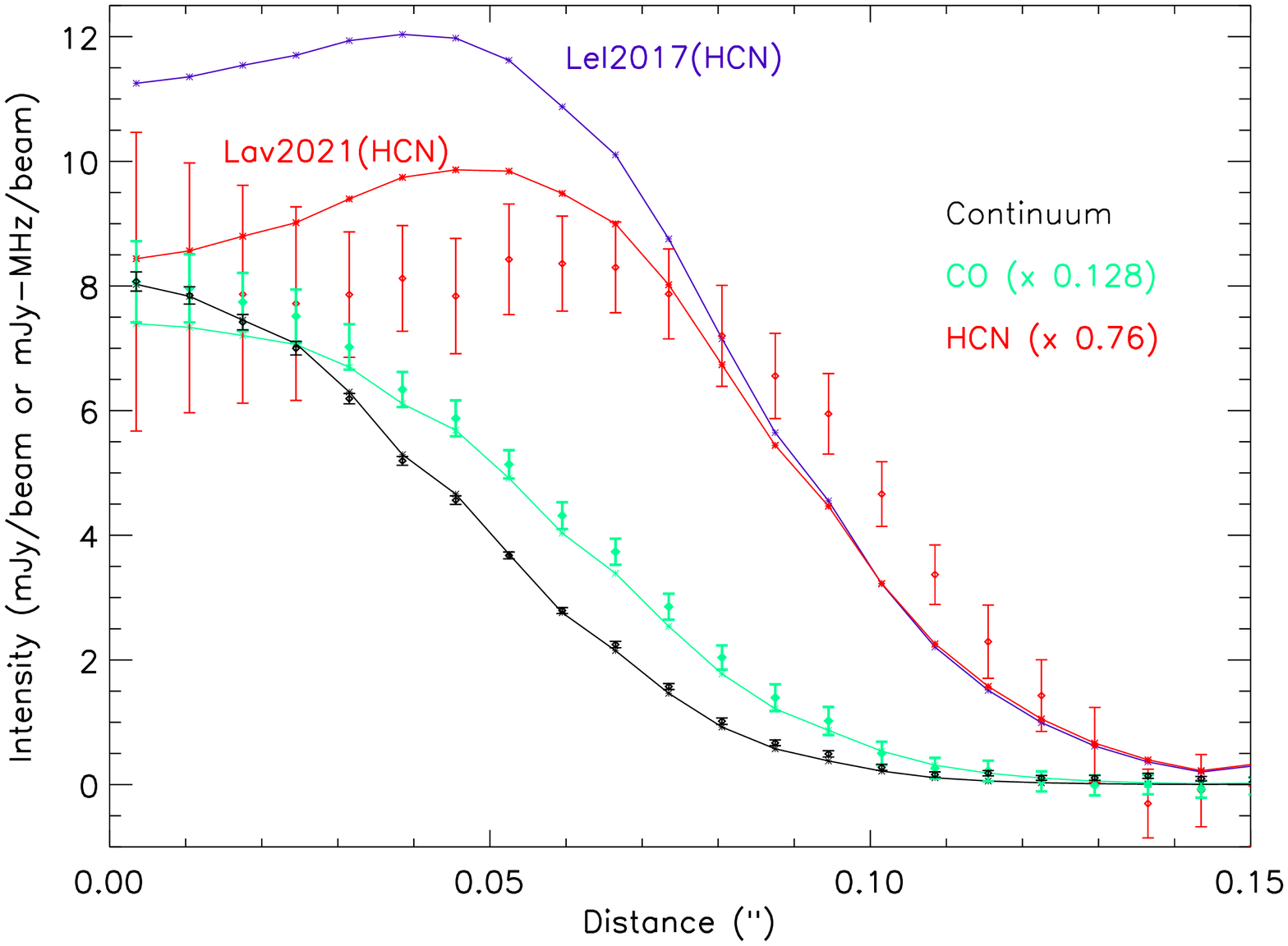}
   \caption{Comparison of the radial profiles of continuum (mJy/beam) and CO and HCN line emission (mJy$\cdot$MHz/beam) from the first panel of Fig.~\ref{Fig_radialprofiles}  with initial models  (see Section~\ref{initial-model}). The thermal profiles and CO and HCN distributions used at this step are shown in the first two panels. For CO, the model
makes use of the Lel2017 (Paper I) thermal profile.  Right panel: comparison between data (points with error bars) and models (continuous lines). Data and models have been scaled vertically as in Fig.~\ref{Fig_radialprofiles} (i.e. multiplicative factors of 0.128 for CO and 0.76 for HCN) for visual convenience. }%
    \label{Fig_radialprofiles_withfits}
   \end{figure*}
To visualize these effects in more details, we averaged spectra in (oversampled) bins of 0.007" in radial distance, and calculated the CO and HCN line areas
for each bin. To estimate error bars on these areas, we generated, for each bin, a series of ``clones" of the observed spectrum, adding random spectral noise
at the relevant level, and calculating the line area for each of these clones; the rms dispersion in the resulting areas yielded the desired error bar.
The first panel of Fig.~\ref{Fig_radialprofiles} shows the relative radial distributions of the continuum (mJy/beam) and CO line and HCN line emissions (mJy$\cdot$MHz/beam). To ease the comparison, the latter are multiplied by constant factors (0.13 and 0.79 respectively) so that all profiles coincide in the first bin centered at a radial distance r~=0.0035". The CO emission is similar in shape with the continuum, though clearly broader. On the other
hand, the radial distribution of the HCN emission is markedly different, peaking slightly beyond Pluto's limb. Second, taking advantage from Pluto's polar axis' roughly SW-NE orientation in the plane of the sky (polar angle PA = 223\dg\ from celestial north), we split radially-averaged profiles twice in two groups,
forming four ``quadrants" (see Fig.~\ref{Fig_COHCN_maps}). Specifically, the RA~=~--DEC line (i.e., from sky SE to NW)
divides the map between ``low Pluto latitudes" (above the line, quadrants 1+2)" and ``high (Northern) latitudes" (below the line, quadrants 3+4), while the RA~=~+DEC line (from sky SW to NE)
divides the map between ``morning" (to the right of the line, quadrants 1+3) and ``evening" (to the left, quadrants 2+4).

While the continuum emission profile does not show any discernible variability between the different azimuth groups (upper right panel of Fig.~\ref{Fig_radialprofiles}), the
enhanced CO line emission at high vs low latitudes is apparent, at the 2$\sigma$ level over radial distances of 0.03-0.07" (lower left panel). For HCN (lower right), the opposite effect is seen (enhanced low latitude emission), possibly combined with a morning vs evening asymmetry, due to Northern mid-latitudes (see Fig.~\ref{Fig_COHCN_maps}).

\section{Modelling and analysis}

\subsection{CO and HCN: modelling}
\label{CO-HCN-model}
To analyze the data, we used our own radiative transfer code, based on standard non-scattering radiative transfer equations, and fully accounting for the spherical geometry associated with the large extent of Pluto's atmosphere. The model, presented in Paper I (where all model details can be found), was extended to cover the 0-1450 km altitude range by means of 200 layers, with an altitude-varying layer thickness. Essentially, spectral radiances are calculated as a function of distance to Pluto center for a series of (surface-intercepting or limb) lines-of-sight (``pencil-beams''), and then convolved with the synthesized beam, in order to calculate spectral fluxes (Jy/beam). Unlike in Paper I, however, where only the flux in the central beam (FWHM$\sim$0.31") was modelled, we implemented the possibility to perform the convolution simultaneously at all spatial points on the observational grid. To minimize computation time, spectral radiances as a function of line-of-sight are calculated once and stored, and spatial convolution is then performed on the entire grid by using suitable weights according to the different beam positions (see \citet{lellouch19}).

\subsubsection{Initial fitting attempts of the emission radial profiles}
\label{initial-model}
In a first step, we compared the radial profiles of Fig.~\ref{Fig_radialprofiles} (first panel) to model expectations based on our previous findings on the thermal structure
and the CO and HCN distributions. In this exercise, we used temperature and HCN profiles inferred from New Horizons and/or our previous disk-averaged ALMA data \citep{lellouch17,lavvas21}\footnote{The two cases correspond to somewhat different thermal and HCN profiles. For \citet[][Paper I]{lellouch17}, we refer to the red profile of their Fig. 5 for the thermal profile and to the "physical profile divided by 2" from their Fig. 7 for the HCN mixing profile. For \citet{lavvas21}, the thermal profile is the New Horizons derived profile (solid line from their Fig. 1) and the HCN profile is shown in their Fig. 2, red curve. All these profiles are shown in the top two panels of Fig.~\ref{Fig_radialprofiles_withfits}.}, assuming they hold at any location on Pluto.  A CO mixing ratio of 515 ppm (Paper I)  
and a brightness surface temperature of 33.0 K 
were used. The adopted thermal and abundance profiles, and the resulting modelled radial dependence of the continuum, CO  emission, and HCN emission (integrated over $\pm$5 MHz and $\pm$1 MHz, respectively) are shown in Fig.~\ref{Fig_radialprofiles_withfits}. The continuum profile is very well fit. The CO radial profile is also generally well reproduced, although the modelled CO line areas are slighty too small (by $\sim$5 \%). Finally, for HCN, our calculations do predict that the HCN emission peaks much further
from disk center than the continuum and the CO emission. However, the modelled HCN emission reaches a maximum at $\sim$0.040-0.052" from disk center, while
in the observations, the HCN emission does not decrease before $\sim$0.065'' 
from disk center, i.e. at $\sim$1.3 Pluto radius. This indicates that the HCN distributions derived from
the disk-averaged ALMA 2015 data are not optimum. In addition, at this stage, the spectral information content of the data has not been fully used. This requires an inversion method,
as we now describe.

\subsubsection{Temperature / CO retrievals}
\label{Temper_retriev}

The observed CO line profiles are sensitive to both the CO mole fraction and Pluto's mean temperature profile in the region probed by the beams.
Following the approach of Paper I, we analyzed the CO data using a constrained and regularized inversion method based on algorithms detailed in 
\citet{conrath98} and \citet{rodgers00}. The method involves the initialization of the CO mixing ratio (assumed constant with altitude) and the 
temperature profile to initial (``a priori'') values. Successive comparisons of model spectra to observations permit to update iteratively the CO and/or
temperatures until model spectra match observations. As detailed in Paper I where the mathematical formulation and the information content are presented, 
temperature profiles are (i) constrained to remain close to the a priori at levels where the measurements contain no information (ii) regularized, 
i.e. smoothed to some vertical resolution (usually comparable to the atmospheric scale height) to avoid unphysical oscillations of the solution profiles.
The inversion scheme was applied to the measured fluxes (in Jy or Jy/beam) over $\pm$30 MHz from the CO line core (i.e. 345.766--345.826 GHz). 
Thanks in particular to the quasi-linear (Rayleigh-Jeans) variation of the radiances with temperatures, convergence to solution profiles is achieved quickly. To avoid ``overfitting" and to minimize the appearance of spurious structures (e.g. oscillations) in the retrieved temperatures, the inversion process was limited to a maximum of two iterations, 
and even only one if the rms between the model and the data, calculated over the fitting interval, had decreased by less than 2 \% after the first iteration. 

\begin{figure*}
   \includegraphics[angle=0,width=\textwidth]{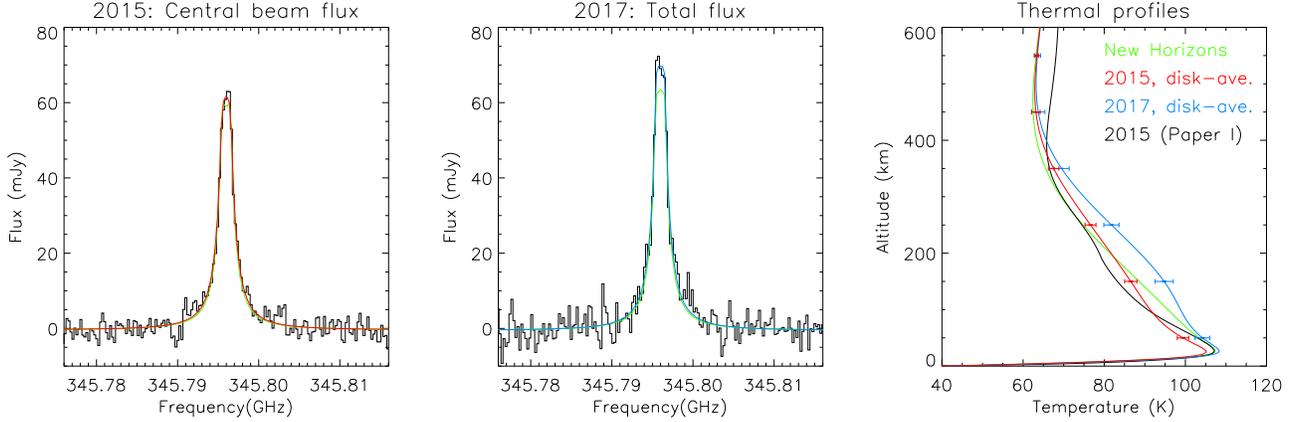}
\caption{Fit of disk-averaged 2015 (central beam position) and 2017 (zero spacing flux) CO spectra. Data have been smoothed to 244 kHz and are shown as histograms.
Models are shown as colored lines with same code as in third panel. The latter shows the retrieved thermal profiles, compared to the a priori (New Horizons, green line) and the temperature profile from Paper I.  }%
    \label{fits_diskavg}
   \end{figure*}

In a first step, we modelled the zero-spacing CO line, i.e. total flux, associated with the combined data from 2017 (last CO panel in Fig.~\ref{zerospacing}).  In doing so,
we initialized the CO mixing ratio to 515 ppm (Paper I). For the a priori temperatures, we initially used 
the New Horizons derived thermal profile \citep{young18}, that combines temperature information from REX below 100 km and line-of-sight abundances from the Alice UV spectrometer above 900 km for N$_2$ and 80 km for CH$_4$. Inversion was performed
simultaneously on the CO mixing ratio and the thermal profile. This returned a CO mixing ratio of 597$\pm$32 ppm. We also remodelled, in the same manner, the CO line from 2015
(central beam position), yielding CO = 571$\pm$23 ppm. The two retrieved CO mixing ratios are, internally, fully consistent, but somewhat larger than the nominal value
from Paper I. The difference stems partly from the different choice of the priori and partly from the fact that the analysis in Paper I had adopted a slightly
too large value for the synthesized beam (0.35", while the actual beam was 0.345" x 0.285"). The spectral fits and retrieved profiles are shown in Fig.~\ref{fits_diskavg}.
The new retrieved disk-averaged profile for 2015 compares well with the New Horizons a priori, with differences mostly within 1-$\sigma$. The disk-integrated 2017 profile based
on the fit of the total flux CO line is warmer than the New Horizons profile, by more than 3 K over 100-400 km (and a maximum of 7 K at 200 km), while the formal, S/N-limited, 
1-$\sigma$ error bar on the retrieval is $\sim$2 K.

In a second step, we inverted the CO spectra at each beam position. 
To ensure comfortable beam oversampling, CO spectra were inverted over a grid of 0.1'' x 0.1'' with 0.01'' sampling (i.e. 21 x 21
spectra to invert). For each beam position, the corresponding spectrum was inverted in terms of a ``local" temperature profile and/or CO mixing ratio, it being understood
that these parameters represent averages over the $\sim$0.06'' beam. 

The inversions were assessed with a number of diagnostics. The fitting quality of the a priori and of the solution were quantified by
$\chi^2$ statistics, both in unreduced form ($\chi^2$ = $\Sigma$ [(model$_i$-obs$_i$)/$\sigma_i$] $^2$ where $i$ refers to the $N$ independent frequencies in the $\pm$30 MHz fitting interval) and in reduced form $S$ = $\sqrt{\chi'^2/N'}$ (in this case, $\chi'^2$ was calculated over the $N'$ frequencies within $\pm$10 MHz, i.e.
the range most sensitive to atmospheric parameters), where a good fit is
associated to $S$ $\sim$ 1. The fitting improvement between the a priori and the solution is characterized by $\Delta$$\chi^2$ = $\chi^2_{solution}$ -- $\chi^2_{a priori}$, namely $\Delta\chi^2$  = -$n^2$ indicates that the solution profile is a better fit than the a priori one at $n$ $\sigma$ confidence level. 

Those retrievals returned solutions at most beam positions within $\sim$0.08'' from disk center, but usually failed at larger distances due to too weak signals. 
CO and thermal profiles retrieved at each beam position were exploited as follows. We first 
extracted CO and thermal profiles as a function of latitude along the polar axis. For this, (RA, DEC) coordinates of points lying on the central meridian from 30\dg S to 90\dg N (and at 60\dg N ``beyond" the north pole, i.e. near the limb), by latitude steps of 30\dg,
were calculated, and the associated CO and temperature values were obtained from interpolation between the three nearest inverted grid points.
We also created ``zonal mean" (of course pertaining to dayside) temperature fields, using two different methods. The first method used only the temperature retrievals at central beam positions that lie within the Pluto disk. Those -- and their uncertainties -- were re-interpolated on a fine (RA, DEC) array (using again the three nearest points of the inversion grid) and the associated thermal profiles were then sorted and averaged by bins of 5\dg\ in latitude, using appropriate weights as per their (altitude-dependent) individual error bars. The second method, termed ``preferred'', 
also made use of retrievals slightly outside (within 0.07'') of the Pluto disk. At each point of the fine (RA, DEC) array, a temperature profile was constructed as a weighted average of all these retrieved profiles, with distance-dependent gaussian beam weighting. These temperature profiles were then again sorted in latitude and weight-averaged according to their error bars. 

    \begin{figure}
   \includegraphics[angle=0,width=\textwidth]{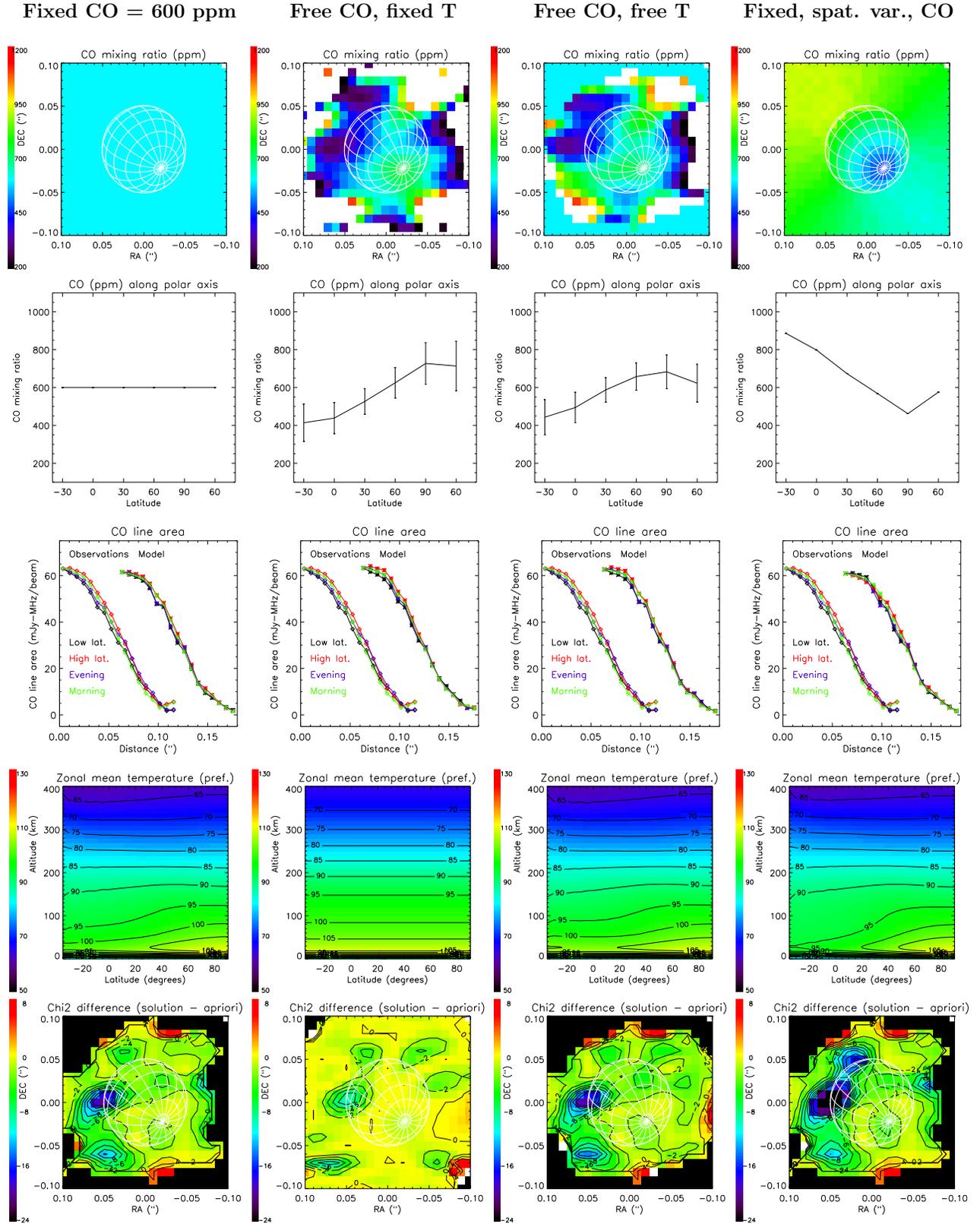}
\caption{Fit results and diagnostics for four variants of the retrievals. For left to right: (1) fixed CO = 600 ppm  (2) fixed thermal profile (3) free CO and thermal profiles (4) fixed, spatially-variable CO. In each case, the five rows show: (i) retrieved CO map. The typical 1$-\sigma$ error bar is 60-100 ppm (ii) CO abundances retrieved along polar the axis, along with their error bars (iii) comparison of observed radial profiles
of the CO line area with model solutions (shifted by 0.06" for better visibility) (iv) retrieved temperature field in zonal mean (``preferred method") (v) fitting improvement between a priori and solution ($\Delta$$\chi^2$).
Beam positions for which the inversion process did not return any solution due to too faint signals are not shown. Additional diagnostics for the first variant are shown in
Fig.~\ref{Fig_retrievals}.}
    \label{Fig_CO_solution_maps}
   \end{figure}

We performed several variants of the inversion, as illustrated in Fig.~\ref{Fig_CO_solution_maps}. In all cases, a 600 ppm CO mixing ratio and the disk-integrated temperature profile retrieved previously were used as a priori. The variants, shown from left to right in Fig.~\ref{Fig_CO_solution_maps}, include the following cases: (1) fixed, uniform, CO = 600 ppm, free thermal profile (2) fixed thermal profile, free CO (3) free CO and thermal profiles (4) fixed, but spatially variable CO, free thermal profiles. This latter scenario
is further described below. For each case,
the different panels in Fig.~\ref{Fig_CO_solution_maps} show, from top to bottom: (i) the retrieved (or assumed, if fixed) CO map (ii) the CO profile along the polar axis (iii) the comparison of the observed radial profiles
of the CO line area (see Fig.~\ref{Fig_radialprofiles}) with model solutions (iv)  the retrieved (or fixed) temperature field in zonal mean (in the ``preferred variant'')  (v) the fitting improvement between the a priori and the solution, as characterized by $\Delta$$\chi^2$.

These retrievals reveal a few noteworthy aspects. First, the free-CO retrievals (either with fixed or free temperatures, second and third columns in Fig.~\ref{Fig_CO_solution_maps})
tend to indicate a CO mixing ratio that increases from low to high northern latitudes. The latter is however barely significant beyond 1-$\sigma$ error bars in the case
of free CO, free temperature retrievals. The free CO, fixed temperature, retrievals (second column) would point to somewhat more marked CO variations, but we note
that it is the one associated with the smallest ``gain" in $\chi^2$ (last line in Fig.\ref{Fig_CO_solution_maps}). Furthermore, a CO increase from the
Equator to the North Pole is not expected. Early post New Horizons general circulation models \citep{forget17} found that
even though sources of atmospheric CO (i.e. surface CO ice, mostly in Sputnik Planitia, e.g. \citet{schmitt17}) are not
widespread on Pluto, the vertical and horizontal uniformity of gaseous CO is expected given its long lifetime against photolysis compared
to atmospheric mixing times. A different situation is however suggested by
recent GCM simulations of Pluto's atmosphere that take into account topography and ice distribution data from New Horizons (Bertrand et al., 2020) and explore three ``climate scenarios” \citep[see Fig. 1 of][]{bertrand20}. These simulations show that the meridional circulation is dominated by a North-to-South N$_2$ (summer to winter) sublimation flow, induced by the presence of N$_2$ ice deposits within Sputnik Planitia and reinforced by mid-latitude deposits in both hemispheres. In the model, the sublimation flow efficiently transports trace gases, including CO, to the southern hemisphere, resulting in larger CO amounts there, with a pole-to-pole contrast by a factor $\sim$2 and a pole-to-equator contrast of $\sim$1.5 in the northern hemisphere. Although this simulated latitudinal distribution of atmospheric CO remains sensitive to model parameters and needs to be investigated in details, we explored the effect of a CO latitudinal gradient of that sign on the temperature retrievals. For that purpose,
we performed a fixed CO retrieval assuming CO($\lambda$) = 800--4$\lambda$ ppm, where $\lambda$ is the latitude in degrees. This case is termed ``fixed, spatially variable CO''.
To specify the CO for beam positions outside of the disk, we used again interpolation between the CO values at the three nearest points in the disk; the resulting injected
CO distribution is shown in the top right panel of Fig.~\ref{Fig_CO_solution_maps}. Interestingly, all three scenarios in which the thermal profile is not kept fixed
indicate a ``hot spot" in the Northern polar atmosphere near 30 km, with zonal mean temperatures elevated by 5-10 K over those at low latitudes (fourth line in 
Fig.\ref{Fig_CO_solution_maps}).


     \begin{figure*}
\includegraphics[angle=90,width=\textwidth]{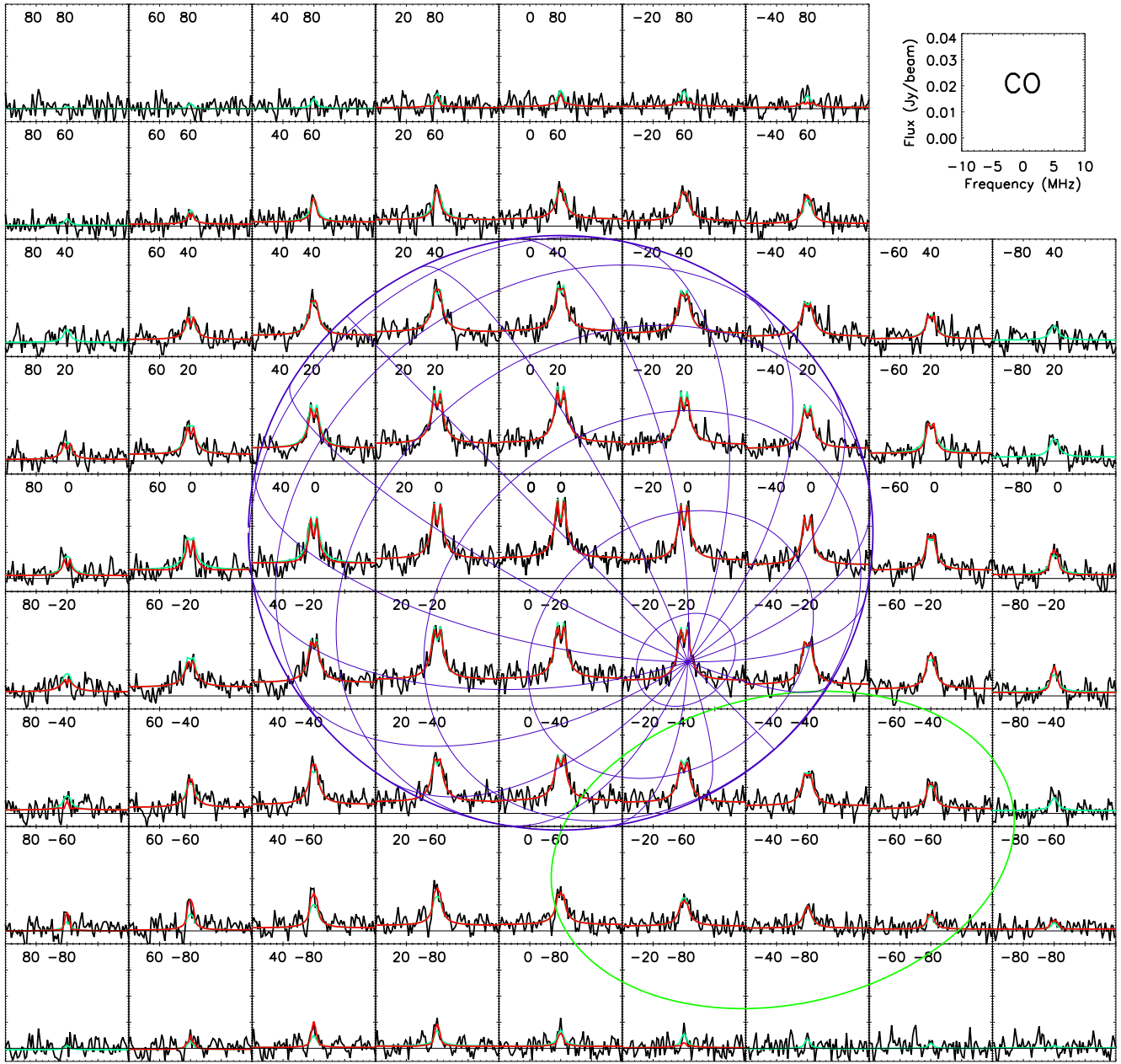}
    \caption{Pluto CO(3-2) spectral map, displayed on a 0.02" grid. Numbers within the panels
give the (RA, DEC) offsets (in milliarcseconds) from disk center. Note the absorption core for beam positions within the disk, directly demonstrating the existence of Pluto's mesosphere. Best-fit models, shown in red, are obtained from temperature inversion, using a fixed CO mixing ratio of 600 ppm. Lacking red curves at a few grid points indicate beam positions where the inversion process did not return any solution due to too faint signals.  Green lines in each panel show synthesized line profiles using the a priori temperature profile. 
The associated retrieved temperature profiles and other parameters are shown in Fig.~\ref{Fig_retrievals}.  The Pluto globe and the synthesized beam are shown.  }%
    \label{Fig_CO_fits}
   \end{figure*}

    \begin{figure*}
     \includegraphics[angle=0,width=18cm]{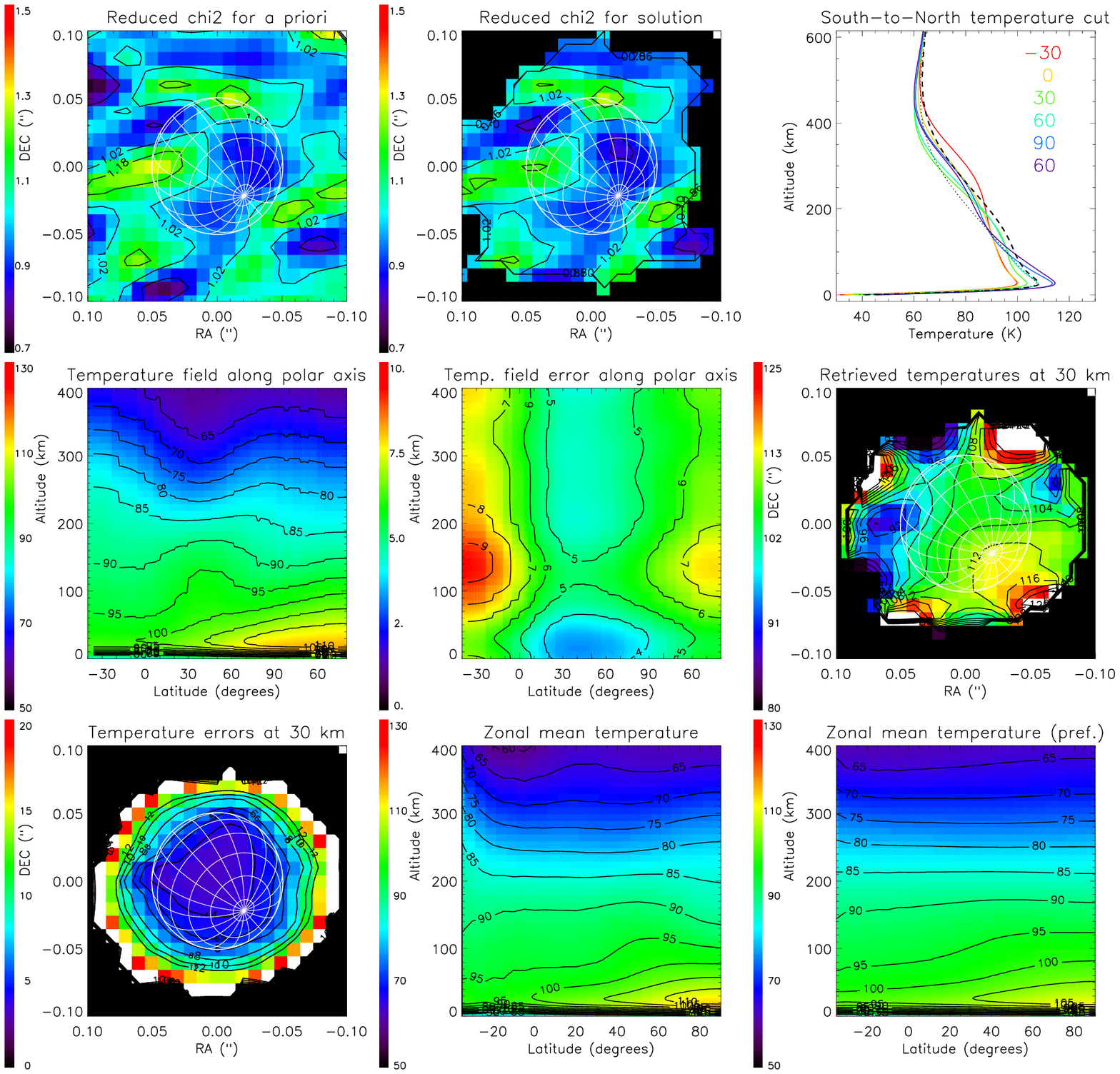}
   \caption{Temperature retrievals for a fixed CO mixing ratio of 600 ppm. For left to right and top to bottom. (a) and b) Reduced $\chi^2$ per point over $\pm$10 MHz from lines for (a) a priori and (b) solution profiles. c)
 Temperature profiles re-interpolated (see text) on the Pluto polar
axis at different latitudes from 30\dg S to 90\dg N and beyond (the 60\dg N purple curve refers to the near-limb 60\dg N point), compared to the initial a priori profile (dashed line)
and the 2015 thermal profile from Paper I (dashed-dotted line); d) and e) latitude-altitude
temperature field along the polar axis and associated uncertainties; f) and g) Retrieved temperatures at 30 km, and their error bars 
h) and i) latitude-altitude temperature field in zonal mean calculated by two different methods (see text). 
}%
    \label{Fig_retrievals}
   \end{figure*}

Focussing on the more conservative, fixed CO = 600 ppm case, Fig.~\ref{Fig_CO_fits} shows the observed CO(3-2) line map, sampled on a 0.02" grid, along with best model fits.
A new observational result to highlight is the presence of absorption cores for many beam positions, most prominent near disk center and progressively vanishing towards Pluto's limb.
This absorption core, which is not apparent in spatially-unresolved observations (see Fig.~\ref{fits_diskavg}), provides a direct spectral evidence for the negative temperature gradient in Pluto's atmosphere above $\sim$30 km, although the latter can be inferred from inversion of the (disk+limb)-integrated line profile (see Paper I and Fig.~\ref{Fig_CO_fits}). Fig.~\ref{Fig_CO_fits} also
shows (green lines) spectral fits using CO = 600 ppm and the a priori temperature profile. At most beam positions, the fits using the a priori and retrieved thermal profiles are very similar, although subtle differences can be discerned on most near limb points.

Additional diagnostics for this CO-fixed (600 ppm) inversion are shown in Fig.~\ref{Fig_retrievals}. 
There, the first two panels show the reduced $\chi^2$ (i.e. $S$ = $\sqrt{\chi'^2/N'}$) over the $\pm$10 MHz interval, for the a priori and retrieved thermal profiles, 
respectively. While the a priori model already provides a good fit to the data at most grid points, a few East limb points are characterized by $S$ $\sim$ 1.2 in the a priori, which can be reduced to $S$ $\sim$ 1.05 in the solution. 
The third panel 
shows a South-to-North series of temperature profiles along the polar axis, interpolated from the grid of retrievals as described above. 
On this plot, the last latitude point (60\dg N; purple curve) corresponds to the 60\dg\ point near the limb ``beyond'' the North Pole.
The fourth and fifth panels of Fig.~\ref{Fig_retrievals} show, respectively, the temperature field along the polar axis, and the associated formal error bars.
Those error bars (1-$\sigma$) purely reflect the noise in the measurements (they are quantified in the inversion
method; see Equations 23 and 24 in \cite{conrath98}). 
The smaller error bars at mid-Northern latitudes compared to the rest of the planet reflect the maximum amount of signal near disk center. Near the stratopause ($\sim$ 30 km), formal error bars on individual profiles are $\sim$3 K near disk center and $\sim$5 K near the low latitude/southern and high northern latitude limbs. Within these error bars, the temperature latitude variations are only $\sim$2-$\sigma$ significant, though the regular increase of the near-stratopause (30 km) temperature from Pluto South to North is noteworthy; the latter can also be seen in the map of the retrieved 30 km temperature and its error bar (sixth and seventh panels of Fig.~\ref{Fig_retrievals}). 
Finally, the last two panels of 
Fig.~\ref{Fig_retrievals} show the zonal mean temperature fields, using the two approaches outlined above. The second method is deemed preferable as using more information
and leading to smoother fields. Signal-to-noise-limited error bars of the zonally averaged temperatures are estimated to be $\sim$2 K over 10\dg N-50\dg N and $\sim$3 K at high poleward latitudes. While the two approaches indicate slightly different temperature fields, 
they both confirm higher summer polar temperatures near
the 30 km stratopause, with an estimated pole-to-equator contrast of 7$\pm$3.5 K. 

While our analysis hints at an elevated stratopause temperature above the summer pole, the comparison of the first two panels of Fig.~\ref{Fig_retrievals}, the maps of $\Delta\chi^2$ (fifth panel of Fig.~\ref{Fig_CO_solution_maps}), and visual inspection of Fig.~\ref{Fig_CO_fits}, admittedly invites to caution. 
Another note of concern may be the fact that atmospheric temperatures at Northern latitudes (50\dg N-70\dg N) are returned to be somewhat higher from the near-limb position ``beyond" the pole than from the near disk-center beam position (third and fourth panels in Fig.~\ref{Fig_retrievals}). 
As temperature variations with local time are unexpected, this might suggest limitations in our analysis of near polar limb data; a conceivable reason would be
that these areas may possibly be associated with a specific surface temperature there. Indeed, while a model with constant surface temperature is sufficient to reproduce the radial variation of the beam-integrated continuum radiances at the spatial resolution of the data (see Fig.~\ref{Fig_radialprofiles_withfits}), such a model is not a physically realistic case, given the known albedo variegation \citep{buratti17} and the observed rotational lightcurves in the thermal infrared \citep{lellouch16}.
We used
an alternate surface temperature description, based on the physical (but still pre- New Horizons) model of \citet{lellouch16}. This model describes Pluto
in terms of three units (N$_2$ ice, CH$_4$ ice, and H$_2$O-tholin, with a distribution shown in Fig. 6 of \citet{lellouch16}), with a thermal inertia of 16 MKS for the non-N$_2$ units. Surface temperatures on the visible hemisphere of Pluto for a sub-Earth longitude of 251\dg, corresponding to one of our CO/HCN observing days (see Table~\ref{obs-tab}), 
were calculated from this model, multiplied by a constant emissivity (0.76) to ensure a 33 K disk-averaged temperature, and convolved by the beam before being injected into the radiative transfer code and inversion tool. This process however led to atmospheric temperature fields that were not significantly different from those inferred with a constant surface temperature model, leading us to believe our result is also robust in this respect. In summary our preferred interpretation of the larger CO line areas
at high vs low latitudes is that they result from enhanced near-stratopause temperatures, by 7$\pm$3.5 K.

\subsection{HCN retrievals}
\subsubsection{Disk-averaged HCN profile}
\label{DiskaveHCN}

We next turned to the study of the HCN vertical profile. As shown in Fig.~\ref{Fig_radialprofiles_withfits}, previous determinations, based on the disk-integrated HCN line
observed in 2015 \citep{lellouch17,lavvas21}, fail to reproduce the radial profile of the line-integrated HCN intensities, which point to an even more extended HCN
emission than predicted by these models. Focussing first on this radial distribution, we performed a parameterized search of the best fit HCN profiles. For this task, we used the retrieved disk-averaged thermal profile for 2017 for all beam positions. In spite of the possible temperature variations in the lower atmosphere we have just reported, this approach is justified by the fact
that HCN mostly probes the upper atmosphere at 600--800 km in its main 354.5047 GHz line. This is higher than the region (up to $\sim$450 km) to which CO is sensitive to; therefore
we have no information on a possible spatial variability of the temperatures in the upper atmosphere. In Paper I, it was also noted that the two satellite lines
of the hyperfine structure, with intrinsic strengths smaller by a factor $\sim$ 45 than the main component, probe the lower atmosphere at 50-150 km. The latter, however,
are not apparent in the spatially-resolved 2017 dataset, due to S/N limitations, and are barely detected in the disk-averaged data (Fig.~\ref{zerospacing}). Information on the upper atmosphere thermal structure is thus best provided by the New Horizons profile \citep{young18}, towards which our CO-derived profiles relax above 400 km.

The HCN profile was parameterized as follows. We started from the q$_{HCN}$(z) profile of \citet[][see above Fig.~\ref{Fig_radialprofiles_withfits}]{lavvas21}. For the lowest
section (z$<$230 km, i.e. from the surface up to the HCN minimum), we used the \citet{lavvas21} profile as such, 
or adjusted by a constant multiplicative factor f$_{low}$. Above 230 km, the HCN profile was described by an increasing power function of altitude up to some
level z$_{top}$ above which q$_{HCN}$(z) is constant at some q$_{top}$ value. Mathematically, q$_{HCN}$ is expressed as:
\begin{equation}
{\rm q}_{HCN}({\rm z}) = {\rm q}_{top} ~ {\rm exp} [~{\rm ln}({\rm q}_{min}/{\rm q}_{top})~ ( \frac{{\rm z}_{top}-{\rm z}}{{\rm z}_{top}-{\rm z}_{min}} )^\alpha~]
\end{equation}
where {\rm q}$_{min}$ is the HCN mixing ratio at {\rm z}$_{min}$ = 230 km in the (possibly rescaled by f$_{low}$) \cite{lavvas21} profile, and $\alpha$ is a parameter characterizing
how this altitude-increasing profile joins with its constant part above z$_{top}$; $\alpha$ $>$ 1 ensures the continuity of the derivative (dq/dz = 0) at z = z$_{top}$.

We ran a grid of models varying f$_{low}$, z$_{top}$, q$_{top}$, and $\alpha$. f$_{low}$ values of 0.01, 0.1, 0.5, 1, 1.5 and 2 were tested; the first two values
are meant to represent cases with a weak or virtually absent lower atmosphere HCN component. As we detail below, the highest values were found to provide unsuitable overall fits.
For the $\alpha$ exponent, we used $\alpha$ = 1.4, 2, 3, 4.5 and 6, which adequately samples a variety of profile
shapes. We explored z$_{top}$ in the range 400-1300 km by steps of 50 km. For each combination of these parameters, the best fit q$_{top}$ value was determined as
the one minimizing the deviation between the model and the observable (i.e. the radial variation of the HCN line-integrated areas), and if the fit was satisfactory,
the range of acceptable q$_{top}$ values was found.

Models were compared to data averaged in 9 bins of radial distance, with a bin size of 0.014", i.e. bins at [0-0.014"], [0.014-0.028"],...,[0.112-0.126"]. 
Because the beam is not circular, it was necessary to calculate synthetic spectra, convolved by the beam, at the precise RA, DEC position of the individual spectra entering each bin and average them for comparison to the data. The metric used to quantify fit quality is therefore 
$\chi^2_{areas, 2017}$ = $\sum_{bins}$ [(A$_{m}$ - A$_{o}$)/$\sigma$$_{A_o}$]$^2$
where A$_{m}$ and A$_{o}$ are the modelled and measured line areas and $\sigma$$_{A_o}$ the error bar on the latter.
 Acceptable fits were defined as those which satisfied a mean deviation, rescaled to the measurement error bar, lower than 1.0 per point in the observable, i.e.
S$_{areas,2017}$ = $\sqrt{\chi^2_{areas,2017}/9}$ $<$ 1.0.

The same parameterized profiles were used to calculate the HCN total (zero-spacing) line in 2017, and the disk-center line in the conditions of the 2015 observation. As recalled
above, that observation was characterized by a beam of $\sim$0.31'', much larger (but not infinitely larger) than Pluto's probed atmosphere, and therefore corresponds almost (but not quite) to the Pluto total flux. For both the 2015 and 2017 lines, the fitting quality of the lines was assessed again from $\chi^2$ analysis, where $\chi^2_{line}$ is now defined as $\chi^2_{line}$ = $\sum_{freq}$ [($\Phi_{m}$ - $\Phi_{o}$)/$\sigma$$_{\Phi_o}$]$^2$, where $\Phi_{m}$ and $\Phi_{o}$ are the modelled and measured (2017 or 2015) spectrum, $\sigma$$_{\Phi_o}$ is the rms noise of the latter 
and the integral runs over the N$_o$ spectra points within $\pm$ 4 MHz from line center; acceptable fits for 2015 were defined by S$_{line,2015}$ = $\sqrt{\chi^2_{line,2015}/N_o}$ $<$ 1.2.

  \begin{figure*}
   \includegraphics[angle=0,width=18.5cm]{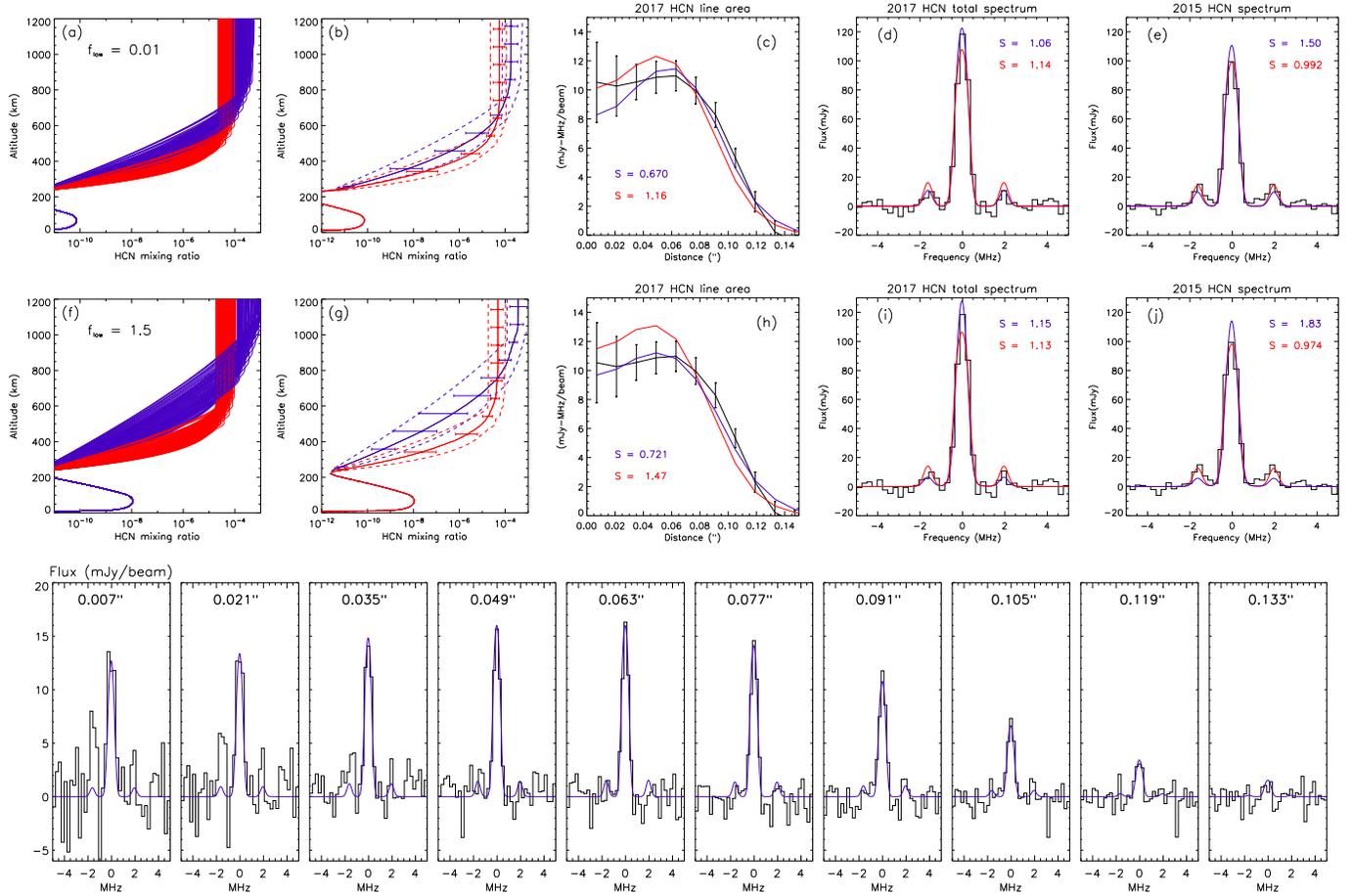}
   \caption{Parameterized fits of the HCN observations. The top two lines show, from left to right; (a and f) family of HCN profiles providing an acceptable fit to the
HCN line area radial distribution observed in 2017 (blue) or to the HCN disk-integrated line observed in 2015 (red). Only a subset of the solution profiles (sampling the full range)
is shown to maintain legibility; (b and g) corresponding median profiles (solid lines), 
extreme profiles (dashed lines), and 16-84 \% percentiles (error bars);
(c and h) Observed (black) HCN line area radial distribution, compared to models using the previous blue and red median HCN profiles. The blue profile provides an optimum
fit; (d and i) Observed (black) HCN line from 2017, after continuum subtraction,  compared to models using these two median HCN profiles ; 
(e and j) Same for the HCN line from 2015. The red profile provides an optimum fit. In panels c, d, e, h, i, j, the reduced $\chi^2_{line}$ values (S$_{areas,2017}$,
S$_{line,2017}$, or S$_{line,2015}$) are indicated with the corresponding color.
  The top line is for model cases with virtually no HCN in the lower atmosphere (f$_{low}$ = 0.01), while  the middle line is for cases with abundant HCN in the lower atmosphere (f$_{low}$ = 1.5). This situation leads to conflicting information provided by the 2015 and 2017 data, with 16-84 \% percentiles being mutually exclusive.
Bottom line: HCN spectra from 2017, averaged by bins of distance (see text), compared
to model calculations using the HCN 2017 median profile for the f$_{low}$ = 0.01 case (solid blue line in panel (b)). }%
    \label{Fig_forward_HCN}
   \end{figure*}

Results are shown in Fig.~\ref{Fig_forward_HCN}. A general result is that while the increase with altitude of HCN, inferred from the 2015 spectrum shape,
is confirmed from the HCN emission radial distribution in the 2017 data, the two datasets do not appear to provide fully consistent results on the HCN
vertical distribution. Specifically, the 2017 observations point to HCN mixing ratios above 200 km increasing more mildly with altitude, but extending
to higher values, than suggested from the 2015 data. This problem is exacerbated when allowance is made for a large HCN component in the lower atmosphere, with domains of solution profiles (16-84~\% percentiles) from 2017 and 2015 data becoming progressively exclusive. The second line of panels in Fig.~\ref{Fig_forward_HCN} illustrates this situation for the
case where f$_{low}$ = 1.5. We also note that regardless of the 2015 data, models that are optimized from the point of view of the radial variation of the line areas provide
a somewhat worse fit to the 2017 total line when a large HCN component is present in the lower atmosphere (blue lines in panels (d) and (i) of Fig.~\ref{Fig_forward_HCN}).
 Based on this, and searching for solutions that minimize differences between 2015 and 2017, we conclude that the maximum rescaling factor of the lower atmosphere HCN component is f$_{low}$ $<$ 0.7, corresponding
to a maximum HCN mixing ratio of 5$\times$10$^{-9}$ in the lower atmosphere (near 65 km). For these cases, the range of HCN column densities\footnote{As in Paper I,
we  account for Pluto's sphericity, we use column densities referred to the surface, i.e. calculated as N$_{col}$ = $\int$ N(z)(1+z/R$_{pl}$)$^2$dz, where z 
is altitude, R$_{pl}$ is Pluto radius and N(z) is the gas concentration at altitude z.} allowed
by the 2017 disk-resolved data and 2015 disk-averaged data are respectively (0.75$\pm$0.25)$\times$10$^{14}$ cm$^{-2}$ and (1.25$\pm$0.45)$\times$10$^{14}$  cm$^{-2}$,
and the profiles fitting both datasets with the above criteria have HCN column densities  in the range (0.90$\pm$0.10)$\times$10$^{14}$ cm$^{-2}$. We can also define the characteristic altitude of the HCN location, z(HCN), as the layer column-density weighted altitude of the HCN profile, ie z(HCN) = ($\int$ z dN$_{col}$) / N$_{col}$. Restricting ourselves to the f$_{low}$ $<$ 0.7 cases, we find that the 2017 (resp. 2015) data indicate z(HCN) = 690$\pm$75 km (resp. 550$\pm$50 km), while the combined solutions
have z(HCN) = 640$\pm$50 km. In the preferred profile for 2017, the HCN line optical depth in vertical viewing is 1.6 at line center and 0.03 in the hyperfine satellite lines.

   \begin{figure*}
   \includegraphics[angle=90,width=\textwidth]{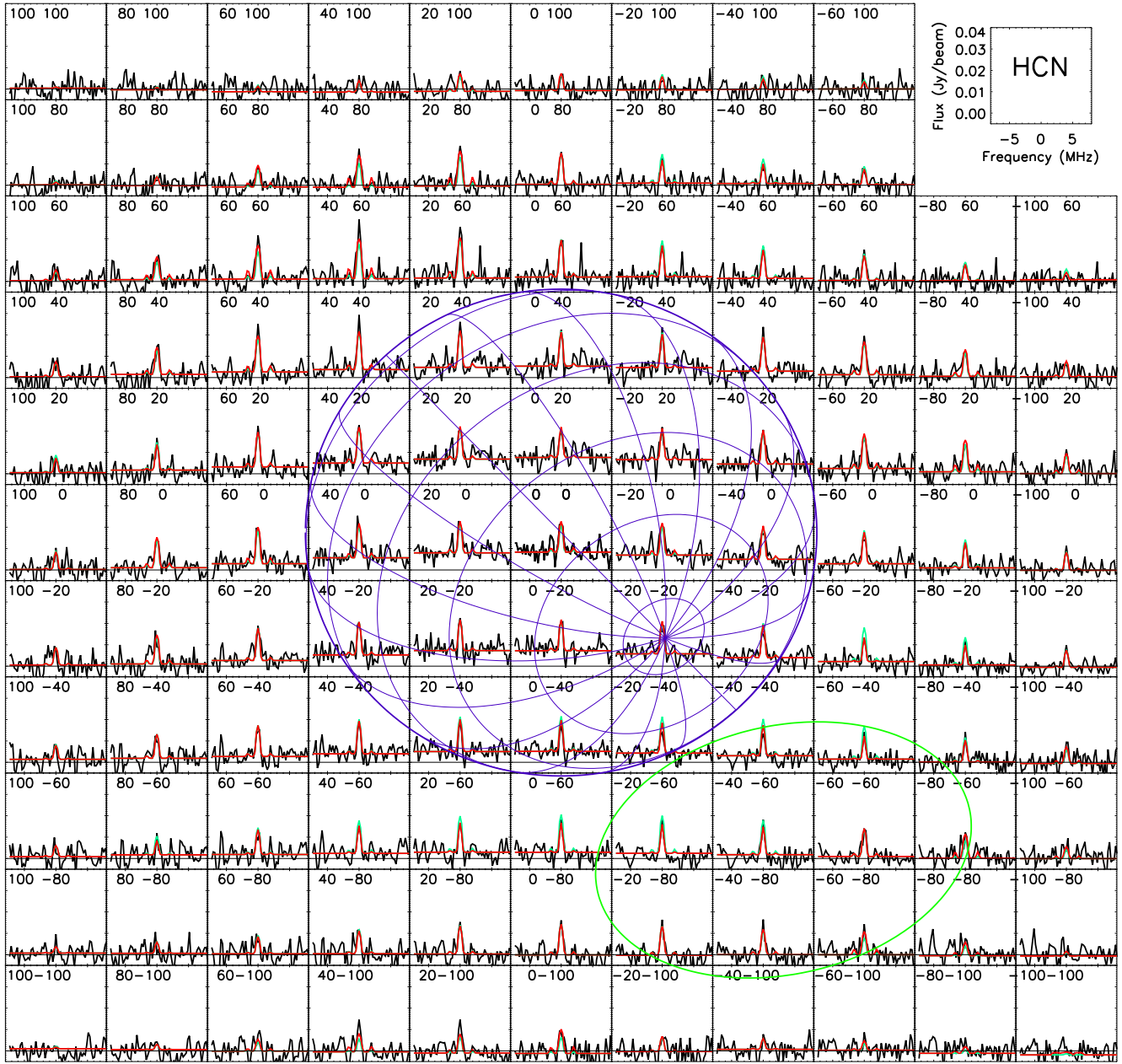}
   \caption{Pluto HCN(4-3) spectral map, displayed on a 0.02" grid. Numbers within the panels
give the (RA, DEC) offsets (in milliarcsec) from disk center. Best-fit models (see details in Fig.~\ref{Fig_invert_HCN_factor}) are shown in red, while green lines show line profiles using the a priori HCN profile derived in Section~\ref{DiskaveHCN}. The Pluto globe and the synthesized beam are shown.  }%
    \label{Fig_HCN_fits}
   \end{figure*}

\subsubsection{Search for HCN spatial variations}
\label{HCNspatvarsearch}
In a second step, we inverted the HCN line profiles at each beam position to search for possible spatial variations of the HCN abundance. In doing so,
we used as initial profile the nominal HCN vertical distribution for 2017 inferred in the previous section under the assumption of a very small HCN abundance in the lower
atmosphere (i.e. f$_{low}$ = 0.01); this distribution is shown as the blue line in panel (b) of Fig.~\ref{Fig_forward_HCN}. When fitting the spectra
at each beam position, we allowed for a single free parameter, namely a constant scaling factor to this profile. 
The reason for this decision is the limited information content of the HCN narrow (minimally pressure-broadened) spectra, preventing us from retrieving a HCN vertical profile from lineshape. As detailed in Paper I, even with high S/N (such as the one achieved on the disk-averaged HCN line in 2015), the number of vertical layers that can be probed independently in terms of the HCN abundance is as low as $\sim$1.5;
essentially, an HCN line profile constrains relatively well the HCN mixing ratio over 600-800 km, and puts a limit on the amount
of HCN in the near-surface ($<$ 100 km) atmosphere, from the appearance and contrast of the HCN satellite lines. This limitation is all the more
true for our 2017 observations, in which the HCN hyperfine structure is not clearly detected. As in the previous section, we considered a single
temperature profile for all beam positions.

%

   \begin{figure*}
   \includegraphics[angle=0,width=\textwidth]{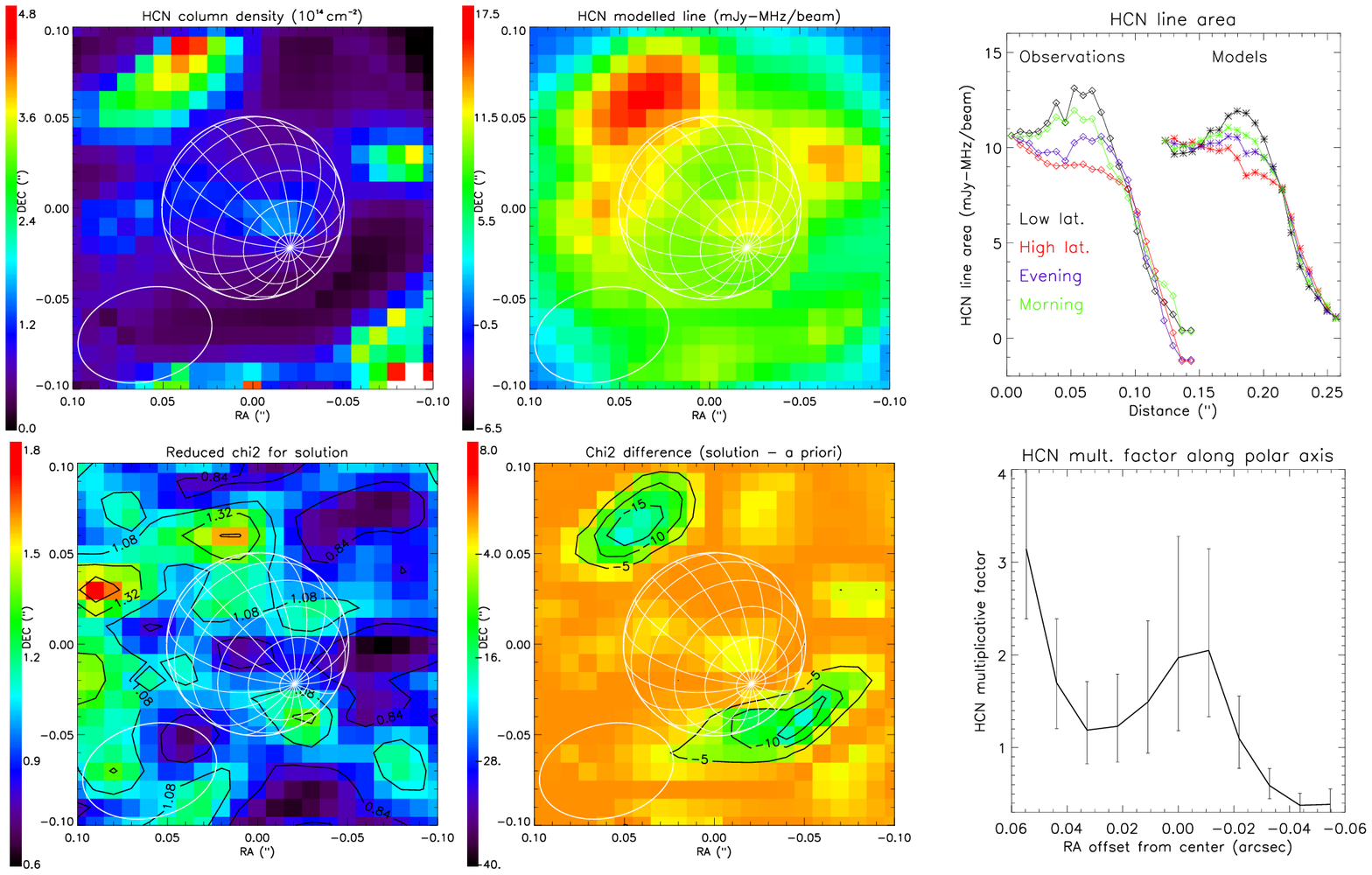}
   \caption{HCN retrieval results, using as a priori the disk-averaged HCN profile for 2017 inferred in Sect. \ref{DiskaveHCN}, and allowing for a multiplicative factor on this profile. From left to right and top to bottom. a) HCN column density at each beam position, color-coded in units of 10$^{14}$ cm$^{-2}$. b) Image of the modelled HCN line emission, integrated over $\pm$1 MHz, to be compared to the fourth panel of Fig.\ref{Fig_COHCN_maps} c)
Compared observed and modelled (shifted by 0.12'' for better visibility) radial profiles of the HCN line emission  d) Reduced $\chi^2$ (S = $\sqrt{\chi^2_{fit} / N}$) over $\pm$4 MHz from line center for best fit  e) $\chi^2$ variation between the a priori and the best fit ($\Delta$$\chi^2$ = $\chi^2_{fit}$ - $\chi^2_{ini}$).
e) HCN multiplicative factor along the polar axis, as a function of RA offset from disk center. 
}%
    \label{Fig_invert_HCN_factor}
   \end{figure*}

Fig.~\ref{Fig_HCN_fits} shows the observed HCN(4-3) line map, sampled on a 0.01" grid, along with model fits, and Fig.~\ref{Fig_invert_HCN_factor} shows fitting results and their diagnostics. Changes in the fits between the initial and retrieved HCN profiles are subtle, and significant differences ($\Delta\chi^2$ = $\chi^2_{fit}$ -- $\chi^2_{ini}$ = -5 to -15,
i.e. 2- to 4-$\sigma$) only occur at/beyond the (sky) North-East and South-West limbs (see fourth panel in Fig.~\ref{Fig_invert_HCN_factor}). Best fits indicate a larger HCN scaling factor
(or equivalently HCN column density) over the North-East (i.e. low latitudes) vs South West (i.e. northern polar latitudes) limbs, by a factor $\sim$4 (first and fifth panels of 
Fig.~\ref{Fig_invert_HCN_factor}) but error bars are large -- formally, the retrieval algorithm returns a typical 1$\sigma$ error bar of $\sim$40 \% on individual spectra. 
Fig.~\ref{Fig_invert_HCN_factor} also shows the comparison of the HCN line area (integrated over $\pm$1 MHz from line center) between observations and retrievals,
in the form of (i) a map of the modelled line areas (third panel, to be compared to the fourth panel of Fig.~\ref{Fig_COHCN_maps}) (ii) the modelled vs observed radial distribution of the HCN areas in the morning, evening, low-latitude and high-latitudes ``quadrants" introduced in Sect.~\ref{COandHCNdata} (see Fig.~\ref{Fig_radialprofiles}).
Although the inversion process does qualitatively reproduce larger HCN line areas at low- vs high-latitudes, the magnitude of
the modelled contrast is somewhat too low; in fact the HCN line areas and contrasts at/beyond the low-latitude (sky North-East) limbs are somewhat underpredicted by the model (this can be seen e.g. in Fig.~\ref{Fig_HCN_fits}, at beam positions near (RA, DEC) = (0.04", 0.04"), and the reduced $\chi^2$ between modelled and observed spectra reaches values of
$\sim$1.4 (fourth panel in Fig.~\ref{Fig_invert_HCN_factor}). In spite of these difficulties,  our observations indicate larger 
HCN amounts, by a factor of several, above Pluto's low-latitude vs high-latitude limb. 
Given that the HCN line emission is mostly optically thin and very extended (effective emission radius 1.40 - 1.78 Pluto radius at 0.5 - 0 MHz from line center),
limb views dominantly contribute to the emission.
As high-northern latitudes beyond $\sim$40\dg\ N contribute little
to the HCN signal, we did not attempt to produce zonal-mean latitudinal profiles of HCN as we did for the temperature field.


We explored another avenue, namely that the spatial variability of the HCN fluxes could be related to temperature variations in the upper atmospheric region ($>$ 600 km) probed by the HCN emission.
For that, we performed temperature inversions, similar to those described in Sect.~\ref{Temper_retriev} for CO, but using this time the HCN lines as diagnostic, keeping the
HCN vertical distribution fixed to the nominal profile from Fig.~\ref{Fig_forward_HCN} (second panel, blue line), and using the disk-averaged temperature profile as a priori. 
This approach provided another, ``technical', solution to the azimuthal variations of the HCN line strength, permitting in particular, a somewhat improved
fit of the HCN lines observed beyond the low-latitude (sky North-East) limb. 
However, the thermal profiles returned at these beam positions were found to exhibit a broad ($\sim$400 km FWHM) ``hot layer", with peak temperatures of 90 K at 800 km. This i.e. $\sim$20~K
higher than the ``canonical" 70 K upper atmosphere temperature, and as discussed below grossly inconsistent with inferences from New Horizons / Alice at low 
latitudes. 
Furthermore, and perhaps not surprisingly given the limited 
information content of the HCN lineshapes, extending these HCN-based temperature retrievals to the entire HCN line map was not very successful. While retrievals
at/beyond the equatorial limb consistently indicated enhanced temperatures there, retrievals at/beyond the polar limb often returned largely discrepant temperature profiles 
even at nearby beam positions. We therefore did not consider this solution any further.

\section{Detection of hydrogen isocyanide (HNC) and upper limits on other species}
\label{HNC}
\begin{figure}
   \includegraphics[angle=90,width=\columnwidth]{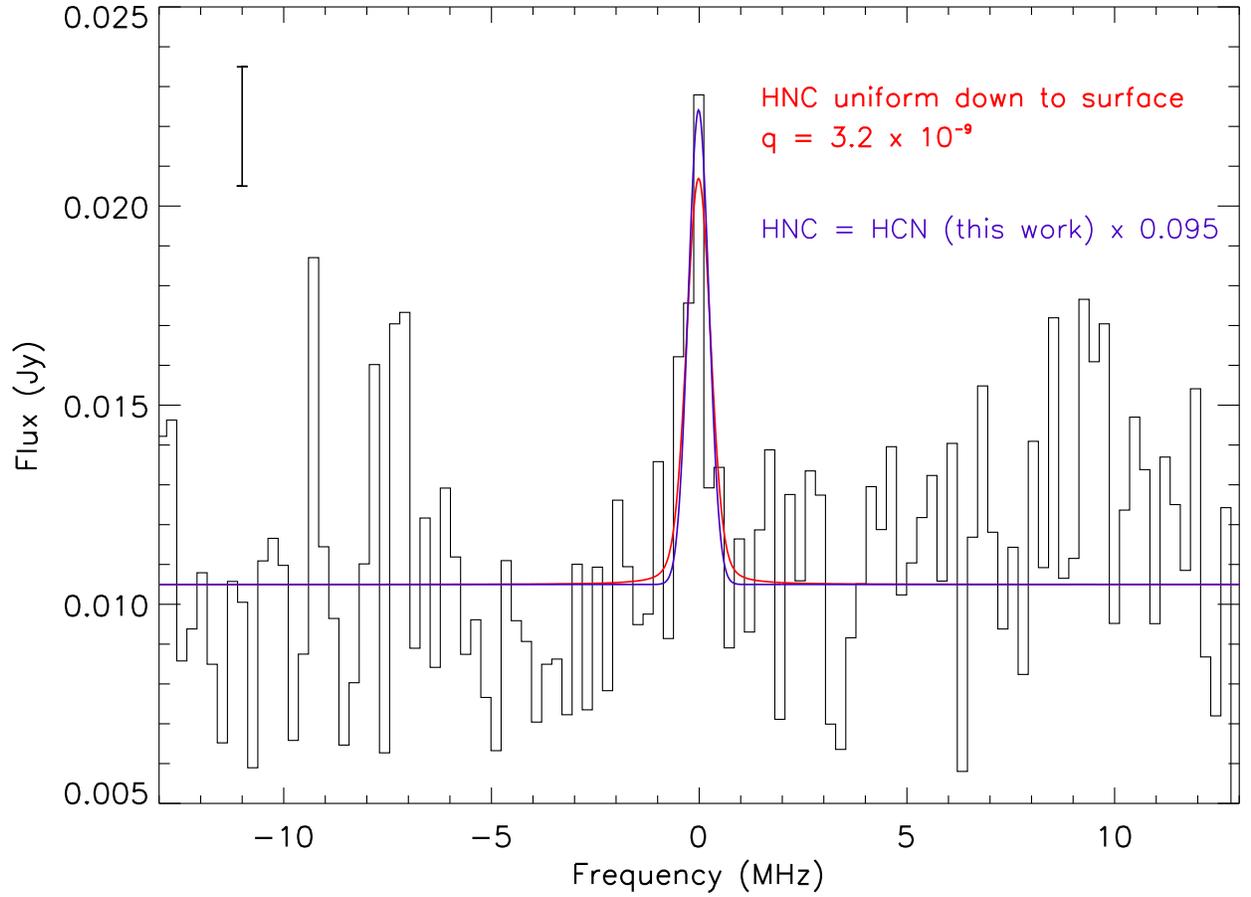}
   \caption{The HNC(3-2) line at 271.981 GHz observed on April 27, 2017 (histograms). Data are smoothed to 0.488 MHz resolution and the vertical error bar indicates the 1-$\sigma$ noise level.  Observations are compared to models with two HNC profiles. (Red): Vertically uniform HNC profile down to Pluto's surface with 3.2$\times$10$^{-9}$ mixing ratio.
Blue: HNC profile obtained by multiplying the best fit HCN profile from this work (see Fig.~\ref{Fig_forward_HCN}) by 0.095 (blue curve in
Fig.~\ref{HNC_profiles}). All other profiles described in Fig.~\ref{HNC_profiles} and Table~\ref{HNC-results} give identical fits. }%
    \label{HNC_detection}
   \end{figure}

Disk-averaged (synthesized beam $\sim$ 0.6'') observations of the HNC(3-2) line at 271.981142 GHz were obtained on April 27, 2017 (see Table~\ref{obs-tab}), using a
122 kHz channel width. Fig.~\ref{HNC_detection} shows the resulting spectrum after smoothing to a resolution of 0.488 MHz. In spite of the
small line contrast (12 mJy at peak, i.e. about 9 times weaker than the disk-integrated HCN(4-3) line, see Fig.~\ref{zerospacing} and Paper I), and
relatively modest S/N (4.0 at the peak), the line is clearly detected, with an area of 6.75$\pm$1.70 mJy$\cdot$MHz. The observation  represents the second
identification of HNC in a planetary atmosphere, after its detection in Titan from {\em Herschel} \citep{moreno11}. The HNC Pluto line is spectrally unresolved, which
is not surprising given that the Doppler-limited linewidth for HNC at 70 K is 0.19 MHz (HWHM) at this frequency; and even if, in an obviously
absurdly extreme situation, the entirety of HNC was confined to a near-surface layer (12 $\mu$bar, 40 K), the associated collisional linewidth would be only 
$\sim$ 2.0 MHz. Therefore, it can be expected that the HNC detection mostly constrains the HNC column density, and not its vertical distribution.

In an approach similar to \citet{moreno11}, we first considered a series of models with uniform vertical distributions of HNC, i.e. with a constant mixing ratio q$_0$
above a given altitude z$_0$, varying z$_0$ from 1000 km to 0 km by steps of 200 km. Results are summarized in Table~\ref{HNC-results}. For a given assumed value of z$_0$,  the S/N-limited precision on the HNC mixing ratio and column is 25-30 \%. As anticipated (and similar to the Titan case), this suite of models determines the HNC column rather well,  $\sim$1$\times$10$^{13}$ cm$^{-2}$ within a factor of $\sim$2, while the HNC mixing ratios associated to these models vary by five orders of magnitude. 
In Titan, the observed HNC linewidths preclude significant amounts of HNC to be present below 400 km \footnote{Subsequent
ALMA observations resolving Titan's limb indicate that Titan's HNC is actually restricted to the thermosphere above 870 km \citep{lellouch19}.}. 
Here in contrast, none of the step-wise models, including the one with HNC down to the surface itself, can be excluded by the data. The latter model, shown in red in Fig.~\ref{HNC_detection}, is slightly different from all others, due to the small contribution of Lorentzian wings, but still compatible with the observations within the noise. The difference with Titan
can be understood by remembering that Pluto's surface pressure typically corresponds to that in Titan's atmosphere near 400 km altitude.

\begin{figure}
   \includegraphics[angle=90,width=\columnwidth]{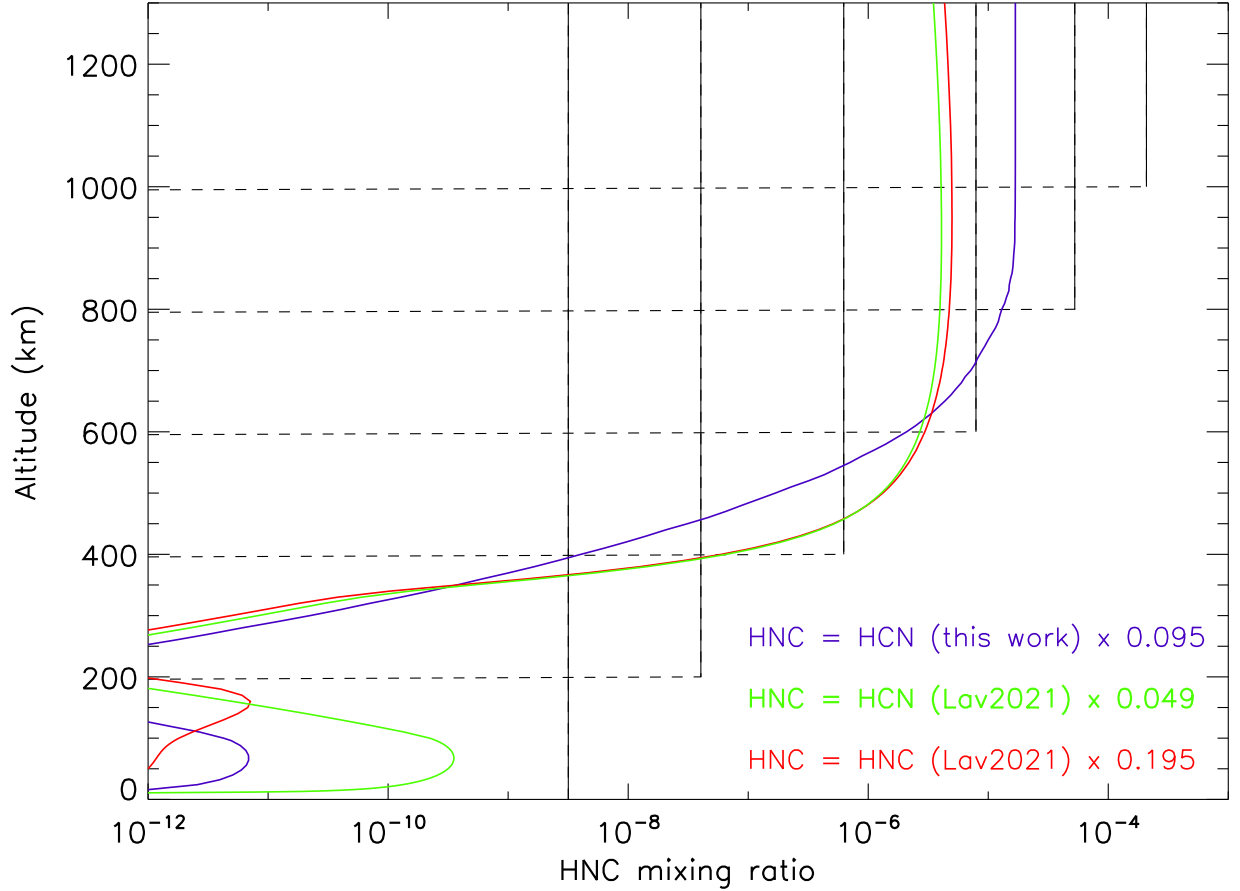}
   \caption{HNC vertical profiles satisfying the HNC observations. Black lines: series of vertically-uniform HNC profiles above some cut-off altitude z$_0$ (1000, 800, 600, 400, 200 and 0 km) satisfying the HNC emission. Colored lines: other solution profiles obtained by rescaling the physically-based profiles of HCN and HNC
from \citet{lavvas21} and the best-fit HCN profile from this work. See text for details.}%
    \label{HNC_profiles}
   \end{figure}

In a second step, and to better constrain the HNC column density, we tested more realistic profiles for HNC. For this we used as ``templates": (a) the HCN profile calculated in the photochemical-microphysical model of \cite{lavvas21}, shown in Fig.~\ref{Fig_radialprofiles_withfits}; (b) the HNC profile from the same model; and (c) the HCN best fit distribution inferred above from our 2017 HCN data (with f$_{low}$ = 0.01, see Fig.~\ref{Fig_forward_HCN}, and Sect.~\ref{DiskaveHCN}). In each case, a scaling factor was applied and determined for best fit
to the HNC spectra (see Table~\ref{HNC-results} and Fig.~\ref{HNC_profiles}). For these three models, fits are identical to those of the step-wise models with z$_0$ $\geq$ 200 km (Fig.~\ref{HNC_detection}). The associated profiles
are shown in Fig.~\ref{HNC_profiles} and described in  Table~\ref{HNC-results}; together they indicate an HNC column density of (7.2$\pm$2.4)$\times$10$^{12}$ cm$^{-2}$.
We regard case (c) as nominal; for this scenario, the HNC/HCN ratio is 0.095, and the HNC(3-2) line center opacity in vertical viewing is equal to 0.14.
Comparison with the \cite{lavvas21} model indicates that the latter overestimates the HNC mixing profile by a factor $\sim$5, and that the model HNC/HCN ratio, about 0.25,
is also too large, by a factor $\sim$2.6.

\begin{table*}
\caption{ \label{HNC-results}HNC mixing ratio and column density for various assumed profiles    }
\centering
\footnotesize
\begin {tabular}{cccccc}
\hline\hline
\noalign{\vspace{3pt}}
Model  & Cut-off &  Rescaled & Mixing & Multiplicative & Column density \\
type & altitude ($\geq$ z$_0$) & profile  & ratio & factor & (cm$^{-2})$ \\

\noalign{\vspace{3pt}}
\hline
\noalign{\vspace{2pt}}
Uniform  & 1000 km & N/A & (2.1$\pm$0.6)$\times$10$^{-4}$ & N/A & (6.4$\pm$1.6)$\times$10$^{12}$ \\
Uniform  & 800 km & N/A & (5.3$\pm$1.5)$\times$10$^{-5}$ & N/A & (6.4$\pm$1.6)$\times$10$^{12}$ \\
Uniform  & 600 km & N/A & (7.9$\pm$2.5)$\times$10$^{-6}$ & N/A & (6.8$\pm$1.9)$\times$10$^{12}$ \\
Uniform  & 400 km & N/A & (6.2$\pm$2.2)$\times$10$^{-7}$ & N/A & (8.0$\pm$2.5)$\times$10$^{12}$  \\
Uniform  & 200 km & N/A & (4.0$\pm$1.4)$\times$10$^{-8}$ & N/A & (1.0$\pm$0.3)$\times$10$^{13}$  \\
Uniform  & 0 km &  N/A  & (3.2$\pm$1.0)$\times$10$^{-9}$ & N/A & (1.6$\pm$0.5)$\times$10$^{13}$ \\
Rescaled (a) & N/A & Lavvas et al. 2021, HCN & N/A & 0.049$\pm$0.014  & (7.5$\pm$2.1)$\times$10$^{12}$ \\ 
Rescaled (b) & N/A & Lavvas et al. 2021, HNC & N/A & 0.195$\pm$0.056  & (7.1$\pm$2.0)$\times$10$^{12}$ \\ 
Rescaled (c) & N/A & This work, HCN & N/A & 0.095$\pm$0.026  & (7.1$\pm$2.0)$\times$10$^{12}$  \\
\noalign{\vspace{2pt}}
\hline\hline
\end{tabular}
\end{table*}

\begin{figure}
   \includegraphics[angle=90,width=\columnwidth]{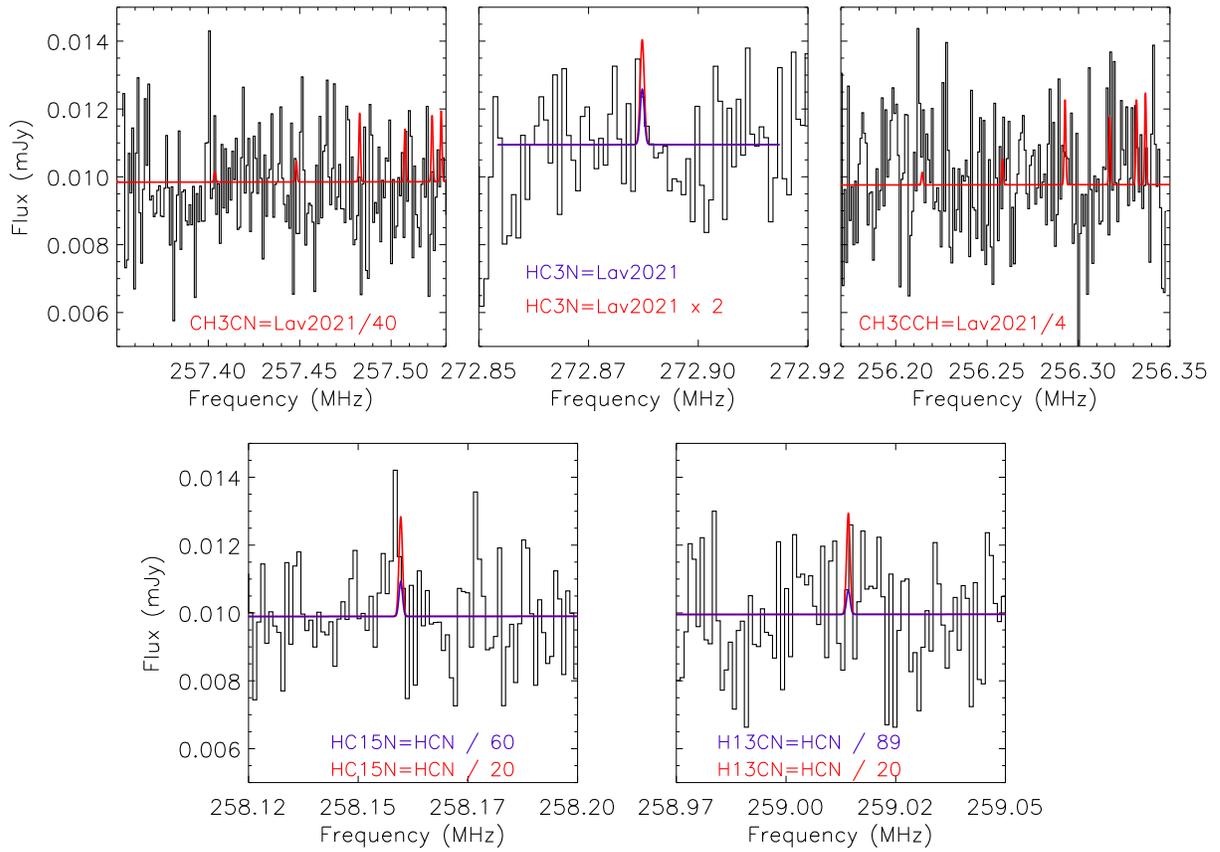}
   \caption{Observed spectra in the region of (i) CH$_3$CN J = 13 multiplet (ii) the HC$_3$N(30-29) line (iii) the CH$_3$CCH J = 14 multiplet (iv)
the H$^{13}$CN(3-2) line (v) the HC$^{15}$N(3-2) line, compared to synthetic spectra. See text for details.
}%
    \label{upper_limits}
   \end{figure}

As described in Sect.~\ref{sec:obs}, the HNC observational set-up also permitted to search for CH$_3$CCH, CH$_3$CN, HC$_3$N, H$^{13}$CN and HC$^{15}$N, none of which were
detected. The corresponding spectra, smoothed to 0.976 MHz resolution, are shown in Fig.~\ref{upper_limits} along with models permitting us to estimate the associated upper limits. For H$^{13}$CN and HC$^{15}$N, we used the HCN profile determined in this study (Sect.~\ref{DiskaveHCN}), rescaled by constant factors. As shown in Fig.~\ref{upper_limits}, we can only
determine upper limits of 1/20 for both H$^{13}$CN / HCN and HC$^{15}$N / HCN. These results are not particularly useful. The $^{13}$C / $^{12}$C
ratio is ubiquitously equal to $\sim$ 1/90 in the Solar System. Regarding  nitrogen, Solar System $^{15}$N/$^{14}$N values vary widely, but even the considerably
enhanced Titan HC$^{15}$N/HC$^{14}$N ratio, that likely results from the combination of a large preferential escape of $^{14}$N over
the evolution of Titan \citep{niemann05} and of a photolysis fractionation effect for HCN \citep{liang07}, is 1/60. A more stringent upper limit 
HC$^{15}$N / HCN $<$ 1/125 was obtained in Paper I.

For CH$_3$CCH, CH$_3$CN, HC$_3$N, we used the photochemical profiles of \citet{lavvas21} as templates and rescaled them by appropriate factors to estimate
upper limits on their column densities. For HC$_3$N, the \citet{lavvas21} profile is consistent with the lack of detection, and we determine an
upper limit equal to twice this profile, corresponding to a maximum column density of 3.0$\times$10$^{13}$ cm$^{-2}$. Note that if instead, we use an HC$_3$N
profile that follows our best fit HCN profile, we obtain a HC$_3$N / HCN ratio of $<$ 0.37 and the same upper limit on the HC$_3$N column density.
A slightly more stringent constraint, but less physically-based, was obtained in Paper I (HC$_3$N column density $<$ 2$\times$10$^{13}$ cm$^{-2}$).
 
In contrast, the CH$_3$CCH and CH$_3$CN
abundance profiles in the \citet{lavvas21} model are too large to be consistent with the data, whose sensitivity is enhanced by the fact that multiple transitions
of each species are present in the observed spectral range (typically four lines of comparable strength, decreasing the effective noise level by a factor $\sim$2).
We find that scaling factors of 1/4 and 1/40, for CH$_3$CCH and CH$_3$CN, respectively, need to be applied to the \citet{lavvas21} photochemical profiles to maintain consistency
with the data. The corresponding upper limits on the column densities are 8.5$\times$10$^{14}$ cm$^{-2}$ for CH$_3$CCH and 2.6$\times$10$^{13}$ cm$^{-2}$
for CH$_3$CN. We note that the CH$_3$CCH upper limit is $\sim$ 3 times smaller than the value reported by \citet{steffl20} from New Horizons Alice
(5$\times$10$^{15}$ cm$^{-2}$ two-way in the solar reflected spectrum off the surface, i.e. 2.5$\times$10$^{15}$ cm$^{-2}$ vertical column).
Profiles / upper limits for all species observed or constrained in this work are summarized in Fig.~\ref{all_profiles} and compared to the \citet{lavvas21} model results.

\section{Discussion}

\subsection{Latitudinal temperature variations in the lower atmosphere?}
Previous information on the variability of Pluto's atmosphere thermal structure was limited to the New Horizons REX and Alice measurements at two low-latitude
locations (about $\pm$16\dg\ latitude) each, yielding very similar temperature profiles, exception made of surface boundary effects.
Despite a number of complications and caveats, as discussed above, our results suggest that summer pole stratopause ($\sim$ 30 km) temperatures may be typically $\sim$7 K
warmer than the corresponding low latitude values. Is this physically plausible? A first consideration is that for the conditions of the 2017 observations (subsolar latitude = 53.5\dg), the insolation, averaged over the course of a Pluto rotation, is $\sim$4.25 times larger at the North Pole than at the Equator\footnote{This effect may be enhanced by 
the fact that high Northern latitudes have higher reflectances than equatorial regions in zonal mean \citep{buratti17}, providing further opportunities for solar heating
after backscattering of solar light at the surface, although this aspect is not considered in models.}. A crude estimate of the order of magnitude of the stratopause temperature variation this may induce can be obtained by applying a ``methane thermostat" model approach \citep{yelle89}.
These authors solved the radiative-conductive equilibrium equation, using simplified expressions for the heating and cooling terms, both assumed to be entirely due to methane.
In the ``isothermal" part of the atmosphere (which in practice corresponds to the region at and just above the stratopause), the heating and cooling terms are balanced.
In the \citet{yelle89} model, non-LTE cooling occurs in the 7.7 $\mu$m $\nu_4$ band of CH$_4$, and is proportional to exp(-h$\nu$/kT), where $\nu$ = 1306 cm$^{-1}$
is the central frequency of the $\nu_4$ band. In constrast, the heating term \citep[absorption in the $\nu_3$ band of CH$_4$ in the][model]{yelle89}, is to first order independent of temperature, and of course proportional to insolation $S$. These considerations suggest that to zeroth order, and assuming no spatial variations of the atmospheric composition (methane abundance), the stratopause temperature T$_{str}$ should be related to the insolation by exp(-h$\nu$/kT$_{str}$) $\propto$ $S$. Assuming a mean T$_{str}$ of 107 K, the factor of 4.25 difference in insolation would then lead to a temperature difference $\Delta$T$_{str}$ = 8.8 K between the North Pole and the Equator. The radiative-conductive
model of \citet{yelle89} model was progressively improved by \citet{lellouch94} who noted the possible importance of CO cooling, \citet{strobel96} who showed the need to include heating in the CH$_4$ 1.6 and 2.3 $\mu$m band complexes, and \citet{strobel17}, who additionally included cooling terms due to HCN, C$_2$H$_2$, based on their distributions
from Paper I and from New Horizons results. In that  model, cooling by C$_2$H$_2$ in its 13.7 $\mu$m band actually dominates the cooling rate
at the stratopause, exceeding the CH$_4$ 7.7 $\mu$m cooling by a factor of 7. Applying the same above relation with $\nu$ = 729 cm$^{-1}$ (13.7 $\mu$m) then leads to a pole-to-equator
difference $\Delta$T$_{str}$ $\sim$ 16 K. These simple arguments thus leave room for explaining a positive 7$\pm$3.5 K temperature gradient from low to high Northern latitudes.

The above estimates hold if atmospheric temperatures reflect instantaneous radiative equilibrium at each latitude. Even in the lack of atmospheric dynamics, latitudinal
contrasts may be decreased in amplitude in relation to the non-zero atmospheric radiative time constant $\tau_r$, roughly by a factor $\alpha^2$ / ($\alpha^2$ + $\omega^2$), where
$\alpha$ = 1/$\tau_r$ and $\omega$ = 2$\pi$/248 yr$^{-1}$ is the angular frequency of Pluto's orbit revolution. \citet{strobel96} calculate $\tau_r$, defined as ``the time required to heat or cool the atmosphere to or from its calculated temperature'', to be typically $\tau_r$ = 10-15 terrestrial years. \citet{strobel17} did not provide specific estimates
of $\tau_r$ from their state-of-the-art model. A rough estimate of $\tau_r$ may be obtained as $\tau_r$(z) $\sim$ T(z) / C(z), where C(z) is the cooling rate in K s$^{-1}$. At the 30 km stratopause peak, where T(z) $\sim$ 107 K and  p = 5 $\mu$bar, the volumic cooling rate  $\rho~c_p$~C(z) is $\sim$4.4$\times$10$^{-9}$ erg cm$^{-3}$ s$^{-1}$  \citep[see Fig. 6b of][]{strobel17}, i.e. 2.75$\times$10$^{-8}$ K~ s$^{-1}$ (using $\rho~c_p$ = 0.16 erg cm$^{-3}$ K$^{-1}$). With such numbers, one obtains $\tau_r$ = 120 terrestrial years, i.e. a reduction of the above contrasts by a factor $\sim$10.
However, a more correct estimate for $\tau_r$, defined as the characteristic time for damping of a small perturbation from radiative equilibrium, is given by
$\tau_r$ = 1 / $\frac{\partial C(z)}{\partial T(z)}$, and at least in Titan's case, this proper definition leads to timescales $\sim$10 shorter than the above expression.\citep{flasar14,bezard18}. If this applies to Pluto as well, $\tau_r$ might actually be closer to 10 years, in (perhaps fortuitous to some extent) agreement with the initial
estimate from \citet{strobel96}, in which case the above 9-16 K estimate of the equator-to-pole gradient would not be modified much. Radiative timescales of
order $\sim$9 years can also be inferred from GCM calculations by \citet{forget17}, who find that simulations initiated with two temperature profiles differing by 30 K
lead to temperature differences less than 2 K after $\sim$25 years.

However, this picture is strongly challenged by dynamical effects, as examined in general circulation models (GCM), developed in preparation \citep{toigo15} or in the wake \citep{forget17} of the Pluto New Horizons 
encounter. In these works, three-dimensional fields are calculated on the basis of theoretical equations of meteorology, incorporating all relevant
processes,  including atmospheric dynamics and transport, turbulence, radiative transfer, molecular conduction, as well as phase changes and volatile transport.
These studies find that for the New Horizons encounter epoch, latitudinal temperature contrasts above the few-km thick boundary layer were small, typically within 1 K, at least on the summer hemisphere\footnote{\citet{toigo15} find that over 1989-2006, but not in 2015, temperatures at the winter pole may be up to 4-6 K colder than at the equator; see their Fig. 16.}.
The lack of temperature contrasts in the models is related to the fact that dynamical mixing times associated to the North-to-South
(summer to winter) flow are short compared to radiative timescales. For typical meridional velocities even as low as $\sim$0.1 m/s (see e.g. Fig. 11 of \citet{toigo15} and Fig. 10 of \citet{forget17}), the time for a parcel of air to travel around half the planet in latitude is of order 1.2 terrestrial year, at least an order of magnitude shorter than $\tau_r$. 
Another factor limiting the horizontal temperature gradients in the GCMs is Pluto's slow (6.39 day) rotation, too weak to maintain geostrophic balance \citep{forget21}.


Pluto's upper atmosphere cold (70 K) temperature is well established, both by New Horizons \citep{gladstone16,young18} and from our 2015 observations (Paper I),
but is challenging to explain. Given its measured vertical profile, and despite its $\sim$7 orders
of magnitude supersaturation in the upper atmosphere, HCN cooling appears insufficient (Paper I). \cite{strobel17} invoked H$_2$O cooling, but this requires 
H$_2$O mixing ratios of order 10$^{-5}$-10$^{-6}$ in the 70 K region above 500 km, i.e. an even more extraordinary degree of supersaturation. Given the difficulty of finding alternate
plausible gaseous candidates, \cite{zhang17} proposed that Pluto's thermal structure is predominantly controlled by haze, not gas, heating/cooling properties. Based on various but realistic assumptions on haze optical properties, deemed to be similar to those of Titan's haze -- in particular the fact that haze particles were continuum absorbers throughout the thermal range with very small single scattering albedos -- \cite{zhang17}'s calculations indicate that haze heating and cooling rates might exceed their gas counterparts, at all altitudes,
by typically 2 orders of magnitude. They further find that the associated cooling time for the haze may be of the order of minutes only, even shorter than
the $\sim$10 day particle settling time in the lower atmosphere, and that the overall atmospheric radiative timescale could be of the order of 
a few months to a few terrestrial years. 

If this drastic revision of our understanding of the energy budget of Pluto's atmosphere is valid, the radiative
timescale could be comparable to the meridional circulation redistribution time, providing a plausible scenario to explain our tentative inference
of latitudinal temperature contrasts. We note however that the similarity of Pluto's and Titan's haze has been questioned by \cite{lavvas21}.
Based on photochemical-microphysical modelling, they find that Pluto's haze formation proceeds by condensation 
of various hydrocarbons on HCN ice condensation nuclei and attendant coating. As a consequence, Pluto hazes may include a major ice component,
dominated by C$_4$H$_2$, with much smaller radiative impact on the atmosphere than tholin-like material (because pure hydrocarbons ices absorb only in discrete bands in the infrared). 
Based on the synthesis of Pluto aerosol analogues, the similarity of Titan's and Pluto's hazes was also questioned by \cite{jovanovic20}, who found that the latter can incorporate   oxygen atoms at the few percent level, possibly increasing their absorption at long UV wavelengths \citep{gavilan17}. In contrast, combining New Horizons
and earlier ground-based data, \cite{hillier21} fitted phase curves for Pluto in various colors and found that Pluto's haze optical properties, in particular
the spectral dependence of the single-scattering albedo $\omega_0$ at visible wavelength, are intermediate between those for Titan and Triton, but more similar with the
former, exhibiting a slope of $\omega_0$ over 500-900 nm like on Titan \citep{tomasko08}\footnote{\cite{hillier21}, however, recognize that the limited Pluto dataset
does not allow them to alleviate the degeneracy between haze single-scattering albedo and optical depth.}. 
In any case,
as pointed out by \cite{zhang17}, the ``haze control" scenario implies a large thermal emission from haze in the mid-infrared, that
could exceed that from Pluto's surface shortward of 25 $\mu$m. The scenario is therefore testable by JWST/MIRI. If true, the amplitude of Pluto's surface 
thermal lightcurve should be progressively erased from 25.5 to 18.0 $\mu$m. 

Finally, our inference of large latitudinal contrasts in Pluto's lower atmosphere tends to contrast with what we know about Triton's atmosphere. Although the
Voyager 2 radio-occultation (RSS) profiles were sensitive to the first $\sim$40 km of the atmosphere but did not provide definite temperature profiles there \citep{gurrola95,oliveira21}, UVS measurements of N and N$_2$ \citep{krasno93} over 170-500 km and 450-750 km respectively yielded remarkably similar profiles at ingress (latitude: 26\dg\ N, winter) and egress (44\dg\ S, summer), suggestive of spatially uniform temperatures despite the large difference in insolation at the two occultation points\footnote{ In contrast, the two CH$_4$ profiles, measured over 0-60 km, were identical below 20 km and different above \citep{herbert91}, as a result of preferential photochemical loss of CH$_4$ in the summer.}. This points to rapid redistribution of heat by winds and to a much weaker role of the haze in Triton's atmosphere thermal budget vs Pluto's, consistent with the
higher and bluer single-scattering albedo of Triton's haze vs Pluto's and Titan's \citep{hillier21}.

\subsection{Chemistry}
\subsubsection{Mean HCN profile}
We performed a parametric exploration of HCN vertical profiles to fit the 2015 disk-integrated HCN spectrum and the 2017 disk-resolved data. This yielded
new constraints on the HCN mean vertical distribution in Pluto's atmosphere, although results based on data from 2017 and 2015 do not appear fully consistent. In the upper atmosphere, 
we find an HCN abundance above 800 km altitude of (1.8$^{+1.8}_{-0.8}$)$\times$10$^{-4}$ for 2017 and (0.55$^{+0.40}_{-0.25}$)$\times$10$^{-4}$ for 2015 (16-84 \% percentiles),
overlapping within error bars, and confirming the high degree of supersaturation of HCN 
found in Paper I. As explained in Section~\ref{DiskaveHCN}, the best overall solutions, that minimize the differences between the two datasets, are found when little or no HCN in present in Pluto's lower atmosphere. We determine 
a maximum 5$\times$10$^{-9}$ mixing ratio near 65 km, and a maximum
column density below 230 km of 1.0$\times$10$^{13}$ cm$^{-2}$. 
Profiles with larger HCN amounts in the lower
atmosphere and that match the radial dependence of the HCN line area in the 2017 data fail to reproduce correctly the detailed hyperfine structure seen in the 2015 data (and to some extent the spectral shape of the 2017 total line). At face value, this result is at odds with our previous report (Paper I) that HCN is present near the stratopause at the 10$^{-8}$ -- 10$^{-7}$ level. As this conclusion was based on the marginally non-zero residual flux in-between the main and satellite HCN lines (i.e. at $\pm$1-1.5 MHz from main line center, see Fig.~\ref{Fig_forward_HCN}(e)), and discarding the possibility of a drastic decline of the HCN abundance in Pluto's atmosphere between 2015 and 2017, we consider the new result to be more reliable. 

HCN is produced in Pluto's upper cold atmosphere ($\sim$70 K) but due to very small nucleation rates (either homogeneous or ion-assisted) and slow condensation kinetics,
insignificant condensation losses for gaseous HCN are expected above $\sim$500 km,  explaining its large supersaturation. In the framework of the \citet{lavvas21} model, HCN ice particles eventually form and settle, and are progressively coated by other condensing gases, largely contributing to haze formation. In the lower atmosphere, where temperatures increase to $\sim$107 K, resublimation of HCN (and other gases) may occur, but this process may be inhibited by the heterogeneous coating.  \citet{lavvas21}
tuned the efficiency of the ice resublimation to match the 10$^{-8}$ -- 10$^{-7}$ HCN abundance in the lower atmosphere reported from Paper I.  Our new upper limit (5$\times$10$^{-9}$ at 65 km) is $\sim$5 times smaller than the saturated value at 100 K (and $\sim$50 times smaller than saturation at 107 K), implying that resublimation is not
complete, and lower than modeled by \citet{lavvas21}. We note that regardless of the extent to which haze particles resublimate, there is also a secondary chemical production of HCN in the lower atmosphere in the model, which leads to a HCN mixing ratio of at least $\sim$10$^{-9}$, consistent with the previous upper limit.  

The HCN column density above 230 km is in the range (0.75$\pm$0.25)$\times$10$^{14}$ cm$^{-2}$ for 2017 and (1.25$\pm$0.45)$\times$10$^{14}$ cm$^{-2}$ for 2015.
Not surprisingly, the latter number matches the corresponding HCN column density in the \citet{lavvas21} model (1.3$\times$10$^{14}$ cm$^{-2}$), since the model
was constrained by this observation. Although error bars overlap, the HCN column inferred for 2017 is nominally smaller than in 2015.
Because the HCN atmosphere lifetime is short, about 0.1 years \citep{krasno20a}, it could conceivably show rapid temporal variations. The solar UV
flux at wavelengths smaller than 150 nm has decreased by ~25 \% in 2017 relative to 2015 \footnote{https://lasp.colorado.edu/home/see/ \citep{woods05}}, and the insolation decrease associated with Pluto's heliocentric distance
is 3~\% from 2015 to 2017. Running the \citet{lavvas21} model with the solar and distance conditions of 2017 leads to a small but non-negligible (15 \%) decrease
of the HCN column. 

\begin{figure}
   \includegraphics[angle=90,width=\columnwidth]{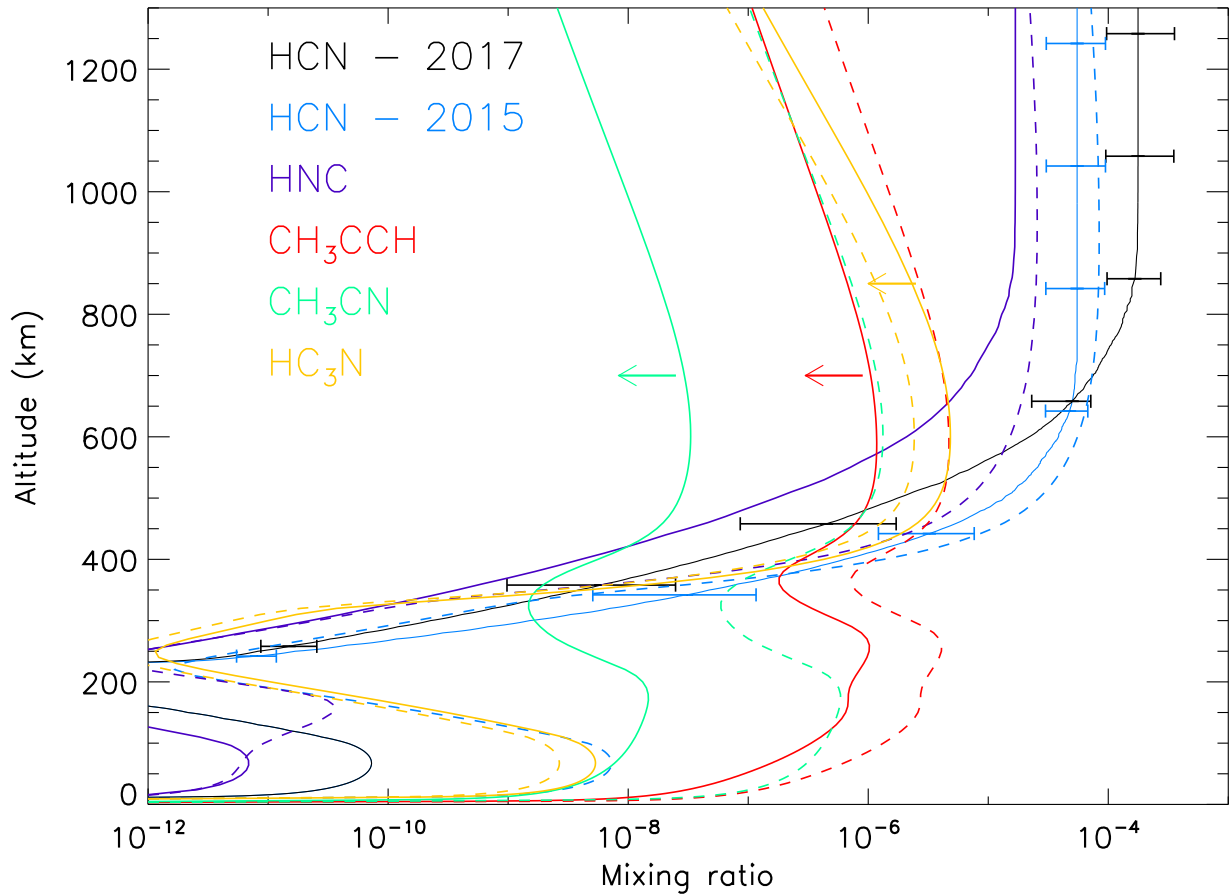}
   \caption{Profiles inferred in this work (solid lines), compared to the model of \citet[][dashed lines]{lavvas21}. For CH$_3$CCH, CH$_3$CN, and HC$_3$N, arrows indicate upper limits
on profiles rescaled from \citet{lavvas21}.}%
    \label{all_profiles}
   \end{figure}

The  \citet{lavvas21} model matches in shape the HCN profile derived for 2015, but the observed profile for 2017 is different
(see Fig.~\ref{all_profiles}), peaking at slightly larger values (e.g., (1.8$^{+1.8}_{-0.8}$)$\times$10$^{-4}$ vs 0.6$\times$10$^{-4}$ in the model for 1100 km altitude), but being less abundant than the model below $\sim$650 km, by a factor up to $\sim$30 at 450 km. Fig.~\ref{Fig_radialprofiles_withfits} directly shows that the \citet{lavvas21} profile does not match the radial variation of the HCN line areas in the 2017 data. Yet, it is not obvious to figure out what modifications in the model could improve
consistency with the observations. \citet{lavvas21} assumed a constant with altitude eddy diffusion coefficient K$_{zz}$ of 10$^3$
cm$^2$s$^{-1}$. However, with an homopause at 12 km \citep{young18}, gas profiles in the middle/upper atmosphere are insensitive to
the precise K$_{zz}$ profile, as the eddy mixing timescale is typically 6 orders of magnitude longer than the diffusive timescale. 
Condensation losses start to affect the HCN profiles below 500 km \citep{lavvas21} and are temperature-dependent. However, as explained above,
temperature information does not extend above 450 km, where the bulk of the HCN lies.

The coupled atmosphere-ionosphere photochemical model of \citet{krasno20a} which, in addition to solar EUV/UV radiation and interplanetary Lyman-$\alpha$, includes galactic cosmic rays (GCR) as a source of photolysis and ionization, 
predicts HCN profiles that are in generally good agreement with our observed profile for 2017.
In this model, condensation dominates the loss of HCN at 150-400 km, while reactions with CH and ions are major losses above 400 km.
The model \citep[Fig. 8 of][]{krasno20a} calculates a HCN mixing ratio of 2$\times$10$^{-4}$ above 800 km, fully matching observations. The lower atmosphere HCN profile, with a 2.4$\times$10$^{-8}$ mixing ratio at 65 km and a 2.5$\times$10$^{13}$ cm$^{-2}$ column below 230 km (V. Krasnopolsky, personal communication), matched the Paper I results (HCN = 10$^{-8}$-10$^{-7}$ below 100 km) but is somewhat too HCN-rich compared to the new observational results for 2017 ($<$ 5$\times$10$^{-9}$ and $<$ 2.5$\times$10$^{13}$ cm$^{-2}$, respectively). Besides the inclusion of GCR, which provide a secondary source of ionization below 300 km and peaking at the surface, 
this model differs from \citet{lavvas21} from a slightly higher eddy diffusion coefficient K$_{zz}$ = 3$\times$10$^4$
cm$^2$s$^{-1}$ (but see the above comment on the expected general insensitivity to K$_{zz}$) and from the treatment of gas condensation on the haze and the surface, parameterized
globally in \citet{krasno20a} by a ``sticking coefficient" $\gamma$ (0.01 for HCN and 0.002 for hydrocarbons).  Similarly, from a neutral-only model,
\citet{wong17} examined the sensitivity of the HCN mixing profile to the sticking coefficients from their photochemical model (with K$_{zz}$ = 1$\times$10$^3$ cm$^2$s$^{-1}$) and also concluded to $\gamma$ = 0.01. 
Rather different results were found by \citet{luspay17} and \citet{mandt17} from their own
ion-neutral model, who used a much higher  K$_{zz}$ = (1--3)$\times$10$^6$ cm$^2$s$^{-1}$ \citep[based on the preliminary CH$_4$ and N$_2$ densities from][]{gladstone16}. They modelled separately the processes of gas condensation and sticking onto aerosols, the latter being required to limit the HCN abundance in the lower atmosphere,
finding that sticking and condensation coefficients of 7$\times$10$^{-4}$ and of 4.5$\times$10$^{-3}$ respectively are
required to provide reasonable agreement with observations from Paper I. A further finding of \citet{mandt17} was that the non-detection of HC$^{15}$N from ALMA and
associated upper limit (HC$^{15}$N / HC$^{14}$N $<$ 1/ 125, Paper I) implies that condensation and aerosol trapping are (unexpectedly) much more efficient for HC$^{15}$N compared to HC$^{14}$N. Although not fully consistent in their approaches and results, these studies illustrate how the determination of the HCN profile can constrain its condensation and adsorption on the aerosol processes.


\subsubsection{HCN spatial variations?}
The HCN(4-3) map (Figs.~\ref{Fig_COHCN_maps} and \ref{Fig_HCN_fits}
) show spatial variations, with stronger HCN emission above
the low-latitude limb. If assigned to compositional heterogeneity, they tentatively indicate variations of the HCN mixing ratio by factors of several along the polar axis,
but the fits are not fully satisfactory (reduced $\chi^2$ per point over $\pm$4 MHz up to $\sim$ 1.4), due to the difficulty for models to reproduce the large line contrasts at low latitudes (see Figs.~\ref{Fig_HCN_fits} and \ref{Fig_invert_HCN_factor}).

An explanation for enhanced HCN at low latitudes does not seem straightforward, given that the HCN abundance is directly related to insolation, which, as discussed
previously, is lower at low vs high Pluto latitudes for the current subsolar latitude conditions\footnote{Although only 1D photochemical models for Pluto are available for now, an analogy may be found with Titan's thermosphere, for which a pseudo-3D model has been developed by \citet{delahaye08} and applied to the latitudes (38.8\dg N and 73.7\dg N) conditions of the Cassini T$_A$ and T$_5$ encounters in 2004-2005, with a subsolar latitude of 23\dg S. Results \citep[see Fig. 24 of][]{delahaye08} indicate diurnal variations of the HCN mixing ratio by factors of several, with maximum HCN in afternoon (12 h-18h) vs early morning (0h-6h), but also that, as may be expected, the HCN mixing ratio is higher (by about
an order of magnitude) at the latitude experiencing the largest insolation (T$_A$).}. In Titan's stratosphere, most minor compounds are enhanced in the winter
hemisphere \citep[e.g.][and references therein]{vinatier20}, as a result of vertical transport of these species in the descending
branch of the global overturning circulation, that also affects haze and temperatures. On Pluto however, the meridional circulation
is relatively weak \citep{forget17,bertrand20},  especially in the upper atmosphere. 
Therefore, we do not expect downward motion in the winter hemisphere causing an HCN enhancement that could be residually visible above the low latitude limb. 
A possible explanation could be related to the fact that in the \citet{lavvas21} description, the nucleation of HCN is mostly driven by the presence of ions. The latter quickly react to insolation changes. A lower ion population at low latitudes could therefore reduce the loss of HCN and conceivably allow for a higher HCN abundance there.

As indicated in Sect.~\ref{HCNspatvarsearch}, another way to fitting the stronger HCN line contrasts above the low latitude limb would be to invoke
a  $\sim$400 km broad ``hot layer", with peak temperatures of 90 K at 800 km. This value is within error bars of temperatures
derived exclusively from the N$_2$ line-of-sight columns measured by New Horizons/ Alice over 900-1000 km \citep[76$\pm$16 K at ingress and 79$\pm$17 K at egress;][]{young18}. 
It is however grossly inconsistent with the inference of a 65-68 K temperature in this region, obtained by combining Alice and REX results with a physical model for CH$_4$ including mixing, diffusion, photolysis, and escape \citep{young18}. In fact, as already shown in Paper I (see their Fig. 9 and 11), a warm layer produces an incorrect slope of the CH$_4$ 
line-of-sight (LOS) columns (measured at all levels from $\sim$100 to $\sim$1200 km) in the corresponding altitude range. 
A reconciliation would therefore imply that Pluto's upper atmosphere at low latitude experienced a drastic $\sim$20 K cooling from mid-2015 to mid-2017.
The scenario raises further serious difficulties. A warm, 400-km thick layer with peak 90 K temperature would imply a total
conduction below and above the peak of $\sim$1.8$\times$10$^{-3}$ erg cm$^{-2}$s$^{-1}$ (using a thermal conductivity k$_{N_2}$ (erg cm$^{-1}$ K$^{-1}$)= 9.37~T (K) ), which is $\sim$5 times larger than can be sustained
by solar EUV/FUV heating of N$_2$ and CH$_4$. A warmer upper atmosphere at low vs high summer latitudes seems also hard to reconcile with
the larger diurnally-averaged insolation (and even more the instantaneous insolation) at the pole. 
Whether, in the \citet{zhang17} scenario of haze control of Pluto's atmosphere energy budget, the haze asymmetry could lead to increased cooling (hence lower high-altitude temperatures at northern latitudes) would remain to be investigated.  

For the above set of reasons, we tend to discard the ``hot layer" solution for the HCN lines. Still we note that Titan's upper atmosphere also exhibits a large and poorly explained temperature variability. The isothermal equivalent temperature of Titan's thermosphere as measured by Cassini / INMS \citep{snowden13} varies between 112 and 175 K.
Remarkably, the data indicate a general lack of correlation between temperature and latitude, longitude, solar zenith angle, or local solar time, but some correlation with Titan's plasma environment, and possible long-term variations (e.g. a 10 K decrease in $\sim$2007 compared to other years). These observational facts may point to particle precipitation from Saturn's magnetosphere possibly being the most significant heat source, but energetics calculations \citep{snowden14} indicate instead that magnetospheric sources are secondary compared to solar input, and that wave dissipation may be a significant source of heating or cooling in Titan's upper atmosphere. While magnetospheric heating is not relevant, waves are known to be present in Pluto's atmosphere \citep[being most directly visible from the layered structure of hazes;][]{cheng17} and their contribution to the heat budget of Pluto's upper atmosphere should also be assessed. 

In summary, we suggest that the spatial variations of the HCN emission point to enhanced HCN abundances at low vs high latitudes, perhaps combined with some temperature
variations in the upper atmosphere, but recognize
we are at loss to propose a satisfactory explanation for them.

\subsubsection{HNC detection and upper limits on other compounds: comparison to photochemical models}
Fig.~\ref{all_profiles} summarizes our findings on the other minor compounds HNC, CH$_3$CN, CH$_3$CCH and HC$_3$N.  For HNC, the suite of step-wise
models determines a column density of $\sim$1$\times$10$^{13}$ cm$^{-2}$ within a factor of 2, and the best guess HNC profile, rescaled from
our determined HCN distribution, has an HNC column density of (7.1$\pm$2.0)$\times$10$^{12}$ cm$^{-2}$. These numbers are very similar to
the HNC column in Titan, (0.6--1.5)$\times$10$^{13}$~cm$^{-2}$ from \citet{moreno11} and 1.0$\times$10$^{13}$~cm$^{-2}$ in the best fit model
of \citet{lellouch19}. Furthermore, Pluto's atmosphere HNC/HCN ratio, 0.095$\pm$0.026 in our preferred model, is strikingly comparable
to that in the HNC and HCN production region near 1000-1200 km in Titan's atmosphere, both in limb-resolving disk observations of \citet{lellouch19} and in the
photochemical models of \citet{hebrard12}, \citet{dobrijevic16} and \citet{vuitton19}. The first of these models considered only neutral production
of HNC, while the latter included ion-neutral chemistry, with the dissociative recombination of HCNH$^+$, as initially proposed 
by \cite{petrie01}, representing a significant source of HNC. 
In contrast, the \cite{lavvas21} model overestimates the HNC column density by a factor $\sim$5, and its HNC/HCN ratio, is also too large by a factor $\sim$2.6 
A critical reaction driving the HNC/HCN ratio is the isomerisation through reaction with HCNH$^+$ (HNCH$^+$ + HNC $\longrightarrow$ HCN + HCNH$^+$).
The rate for this reaction is not measured, and currently estimated in models. In \cite{lavvas21}, this reaction was simply turned off in order to optimize
the calculated HCN abundance with respect to the 2015 measurements. Including it back in the reaction scheme does decrease (resp. increase) the HNC (resp. HCN) abundance.
The resulting HNC column density, 1.4$\times$10$^{13}$~cm$^{-2}$, is still overpredicted by a factor $\sim$2, but the HNC/HCN ratio becomes 0.11, now consistent with the measured range.
We note that other post New Horizons photochemical models \citep{wong17,krasno20a} did not include HNC, and our detection provides a new constraint.

Our upper limits on CH$_3$CN and CH$_3$CCH are 2.6$\times$10$^{13}$ cm$^{-2}$ and 8.5$\times$10$^{14}$ cm$^{-2}$, respectively 40 times and 4 times smaller than in 
the \citet{lavvas21} model. CH$_3$CN was apparently not included in earlier photochemical models, except in \citet{krasno20a}; CH$_3$CN profiles are not
shown in that paper, but the associated column density is $\sim$1$\times$10$^{13}$ cm$^{-2}$ (V. Krasnopolsky, priv. comm.), in agreement with our determined upper limit.
For CH$_3$CCH, the limit is also lower than or comparable to
predictions by other photochemical models. According to \citet{steffl20}, the model by \cite{wong17} includes a 1.25$\times$10$^{15}$ cm$^{-2}$ CH$_3$CCH column.
\citet{krasno20a} predicted a 9$\times$10$^{14}$ cm$^{-2}$ column, but prompted by the reported detection of CH$_3$CCH by \citet{steffl20} from the
UV Pluto spectrum observed by New Horizons Alice in nadir mode, \citet{krasno20b} revised rate coefficients for three reactions following \citet{vuitton19} and obtained 
a 4.3$\times$10$^{15}$ cm$^{-2}$ CH$_3$CCH column. 

As discussed above, our new upper limit on the HCN mixing ratio in the lower atmosphere indicates a reduced re-sublimation in the lower atmosphere with respect to that assumed in \citet{lavvas21}.
This also must imply lower abundances of CH$_3$CN and CH$_3$CCH. Running the model with the sublimation of the ice particles turned off 
leads to smaller CH$_3$CN and CH$_3$CCH columns by factors of 12 and 24, respectively, compared to the original \citet{lavvas21} results. This brings the modelled
CH$_3$CCH in comfortable consistency with our observational upper limit, but the model continues to overpredict CH$_3$CN by at least a factor of 3.5. Besides possible
uncertainties in the temperature dependence of some reaction rates, a plausible cause for the remaining discrepancy may be 
the lack of a sublimation law for CH$_3$CN in the relevant temperature range \citep[see][]{fray09},
making the extrapolation of vapor pressures measured at 280 K extremely uncertain.

Regardless on the comparison to models, our upper limit on CH$_3$CCH is in fact inconsistent (3 times smaller) with the 2.5$\times$10$^{15}$ cm$^{-2}$ column density inferred by \citet{steffl20}\footnote{These authors mention a 5$\times$10$^{15}$ cm$^{-2}$ in their abstract, but their Figure 6 indicates this number is the two-way column.}. Including CH$_3$CCH does allow one to match an otherwise unexplained absorption feature in Pluto's spectrum at 1540~\AA\, but leaves similar features at 1530~\AA\ and 1570~\AA\ unaccounted \citep{steffl20}.
While other, yet to-be-identified compounds may be invoked to resolve this issue, \citet{steffl20} further mention that their re-examination of the \citet{young18} solar occultation
profiles indicates that a 1540~\AA\ absorption feature is present at tangent altitudes from 100 to 150 km, from which a preliminary reanalysis yields a 1.5$\times$10$^{15}$ cm$^{-2}$ CH$_3$CCH column along the line-of-sight (LOS). This is worrisome because the LOS-derived column should be higher than the nadir column, by a typical factor
$\sqrt{2\pi R/H}$ which equals $\sim$13 for R$\sim$ 1315 km and H~$\sim$~50 km; that is, a 1.5$\times$10$^{15}$ cm$^{-2}$ horizontal column would indicate a $\sim$1$\times$10$^{14}$
cm$^{-2}$ vertical column. Given this and our upper limit, we are led to some skepticism about the CH$_3$CCH detection of \citet{steffl20}.

Finally our HC$_3$N upper limit, 3$\times$10$^{13}$ cm$^{-2}$, not better than in Paper I, is reasonably consistent with model predictions (1.5$\times$10$^{13}$ cm$^{-2}$ from \citet{lavvas21}, 3.9$\times$10$^{13}$ cm$^{-2}$ from \citet{krasno20a} and 2.3$\times$10$^{13}$ cm$^{-2}$ from \citet{krasno20b}. 

\section{Summary}
Building up on our previous detection of CO and HCN in disk-integrated observations, we acquired new ALMA observations of Pluto's atmosphere in 2017, with two main goals : (i) spatially resolving the  CO(3-2) and HCN(4-3) line emission; and (ii) searching for new species, expected from photochemistry and analogy with Titan. Analyzing the data with  
radiative transfer and inversion techniques, we obtained the main following results:
\begin{itemize}
\item The CO(3-2) line map shows subtle spatial variability of the line appearance across the disk, that we tentatively attribute to warmer temperatures in the lower atmosphere
($\sim$30 km) at high northern summer vs low latitudes, with a contrast of 7$\pm$3.5 K contrast at 30 km altitude. If confirmed, these unexpected variations
point to a possible radiative control of Pluto's atmosphere by the haze, with relatively short (terrestrial months/few years) radiative timescales compared to gas-only
timescales (10-15 years).
\item The radial distribution of the HCN(4-3) emission, considerably more extended than its CO(3-2) counterpart, provides a new method to determine the average HCN vertical profile.
The bulk of HCN is located at a mean (column density weighted) altitude of 690$\pm$75 km. Conversely,
the HCN component in the lower atmosphere ($<$200 km) is lower than previously inferred from disk-averaged data, implying that resublimation
of the presumably ice-coated aerosols is partly inhibited.
\item The HCN(4-3) line map shows larger HCN emissions above the low-latitude vs high-latitude limb, which suggests an enhanced HCN abundance there, but our models are only partly successful at reproducing them, and the interpretation is uncertain.
\item The first detection of HNC in Pluto's atmosphere is reported, with a (7.0$\pm$2.1)$\times$10$^{12}$
cm$^{-2}$ column density and a HNC / HCN ratio of 0.095$\pm$0.026, very similar to their values in Titan's atmosphere. 
\item New upper limits on CH$_3$CCH and CH$_3$CN are obtained. The non-detection of CH$_3$CCH seems inconsistent with its reported identification from New Horizons.
\end{itemize}

While some of our results are tentative due to S/N limitations, we note that substantial progress would require investing significantly longer
integration times (typically 36 hours of ALMA instead of 4 hours for a CO / HCN mapping three times deeper than presented here). Obviously, in the lack of another
space mission to the Pluto system in the foreseeable future, further observational characterization of Pluto's time-changing atmosphere will rely on
ground-based or remote-sensing techniques (e.g. stellar occultations, infrared and mm spectroscopy) and facilities, e.g. JWST, VLT/CRIRES+, TMT, E-ELT \citep{buie21}, in addition to ALMA. The observations described here represent the first disk-resolved spectroscopic data of Pluto's atmosphere obtained from Earth.
In the relatively near future (2025-2027), however, thirty-meter class telescopes (TMT, E-ELT)
will gather $\sim$10 times more light than the current most powerful optical telescopes, and equipped
with adaptive optics, will achieve a typical diffraction-limited 10 mas spatial resolution (in H band), providing new insights
in Pluto's climatic system.

\vspace*{2cm}

{\bf Acknowledgements:}\\
This paper is based on ALMA program 2016.1.00426.S. ALMA
is a partnership of ESO (representing its member states), NSF
(USA) and NINS (Japan), together with NRC (Canada), NSC and
ASIAA (Taiwan), and KASI (Republic of Korea), in cooperation with
the Republic of Chile. The Joint ALMA Observatory is operated
by ESO, AUI/NRAO and NAOJ. The National Radio Astronomy
Observatory is a facility of the National Science Foundation operated under cooperative agreement by Associated Universities,
Inc.

\newpage

\label{}

\bibliography{bibliography.bib}




\end{document}